\documentclass[11pt,letterpaper]{article}

\usepackage[letterpaper,margin=1in]{geometry}
\usepackage[T1]{fontenc}
\usepackage[utf8]{inputenc}
\usepackage{lmodern}
\usepackage{microtype}
\usepackage{amsmath,amssymb,amsthm,mathtools}
\usepackage{hyperref}

\hypersetup{
  hidelinks,
  pdfauthor={Xiaohong Cai, Mingyang Zhu},
  pdftitle={Information Complexity of Inner Product and Matrix Multiplication via Shannon-to-Min-Entropy Conversion},
  pdfsubject={arXiv preprint}
}
\allowdisplaybreaks[2]
\newtheorem{theorem}{Theorem}[section]
\newtheorem{lemma}[theorem]{Lemma}
\newtheorem{proposition}[theorem]{Proposition}
\newtheorem{corollary}[theorem]{Corollary}
\theoremstyle{definition}

\theoremstyle{remark}

\newcommand{\E}{\mathbb E}
\newcommand{\F}{\mathbb F}
\newcommand{\Z}{\mathbb Z}
\newcommand{\1}{\mathbf 1}
\newcommand{\Pois}{\operatorname{Pois}}
\newcommand{\IP}{\operatorname{IP}}
\newcommand{\MM}{\operatorname{MM}}
\newcommand{\IC}{\operatorname{IC}}
\newcommand{\supp}{\operatorname{supp}}
\newcommand{\rank}{\operatorname{rank}}

\begin{document}

\hypersetup{pageanchor=false}
\begin{titlepage}
\renewcommand{\thefootnote}{\arabic{footnote}}
\pagestyle{empty}
\thispagestyle{empty}
\centering
\vspace*{0.15in}
{\Large\bfseries Information Complexity of Inner Product and Matrix Multiplication via Shannon-to-Min-Entropy Conversion\par}
\vspace{0.3in}

{\large Xiaohong Cai\footnote[1]{The Chinese University of Hong Kong, Hong Kong SAR, China (e-mail: \href{mailto:cx021@ie.cuhk.edu.hk}{\nolinkurl{cx021@ie.cuhk.edu.hk}}).}\hspace{3em}Mingyang Zhu\footnote[2]{Corresponding author. School of Information and Electronic Engineering, East China Normal University, Shanghai 200241, China (e-mail: \href{mailto:myzhu@ee.ecnu.edu.cn}{\nolinkurl{myzhu@ee.ecnu.edu.cn}}).}\par}
\par\vspace{0.35in}

\begingroup
\leftskip=0.03\textwidth
\rightskip=0.03\textwidth
\parfillskip=0pt plus 1fil
\begin{center}
\textbf{Abstract}
\end{center}
\normalsize
\setlength{\parindent}{1.5em}
\noindent
We determine the information complexity of matrix multiplication over finite fields under independent uniform inputs, up to a universal additive constant. The binary inner-product case resolves the open question of Braverman, Garg, Pankratov, and Weinstein on the optimal dependence of information complexity on the error probability. For inputs in \(\F_2^m\) and error at most \(\varepsilon\in[0,1/2]\), the information complexity with output at Bob lies between \((1-2\varepsilon)m-c_0\) and \((1-2\varepsilon)m\), where \(c_0<4.772\) is a universal constant. Thus, the optimal first-order dependence on error has coefficient \(2\).

For independent uniform matrices \(\boldsymbol X\in\F_q^{t\times m}\) and \(\boldsymbol Y\in\F_q^{s\times m}\), where \(t\le s\), computing \(\boldsymbol X\boldsymbol Y^{\mathsf T}\) with output at Bob and error at most \(\varepsilon\in[0,1]\) has information complexity \(\bigl(1-q^{ts}\varepsilon/(q^{ts}-1)\bigr)^+mt\log_2q\), up to an additive gap of less than \(9.544\) bits. This gap is independent of the matrix dimensions, field size, and error. At zero error, the information complexity is exactly \(mt\log_2q\).

The information lower bounds for both inner product and matrix multiplication follow from a new Shannon-to-min-entropy theorem for convolution on discrete groups, which we prove in this paper. Min-entropy is determined by the largest probability assigned to a single outcome. Let \(Q_1,\ldots,Q_k\) be the finitely supported distributions of independent random variables taking values in a discrete group, and let \(s\) be the largest support size. We show that if \(k\ge\lceil\lfloor\log_2s\rfloor\ln2\rceil\), then the min-entropy of their ordered product is at least the average Shannon entropy of the \(Q_i\) minus \(c_0\) bits. This holds for arbitrary discrete groups, including infinite and noncommutative groups, even when the input distributions differ. The leading coefficient of the logarithmic threshold is optimal.

\par
\endgroup
\vfill
\end{titlepage}

\hypersetup{pageanchor=true}
\pagenumbering{arabic}
\setcounter{page}{1}

\section{Introduction and Main Results}\label{sec:introduction}

How much information can a communication protocol save by allowing errors? We first consider binary inner product: Alice and Bob receive independent uniform vectors \(\boldsymbol X,\boldsymbol Y\in\F_2^m\), and Bob must output \(\IP_m(\boldsymbol X,\boldsymbol Y)=\boldsymbol X^{\mathsf T}\boldsymbol Y\). Alice and Bob may use random bits shared between them (public randomness), as well as private random bits available only to one of them. The transcript \(\Pi\) consists of the shared randomness and all exchanged messages. For a protocol with inputs \((\boldsymbol X,\boldsymbol Y)\) drawn from any joint distribution \(\mu\), its information cost is
\begin{equation}\label{eq:internal-information}
\IC_\mu(\Pi)=I(\boldsymbol X;\Pi\mid\boldsymbol Y)+I(\boldsymbol Y;\Pi\mid\boldsymbol X),
\end{equation}
where \(I(\cdot;\cdot\mid\cdot)\) denotes conditional mutual information. Information complexity is the infimum of this cost over protocols satisfying the specified output and error constraints.

For binary inner product, a simple protocol gives a linear saving. With public probability \(1-2\varepsilon\), Alice sends her entire input and Bob computes the answer. Otherwise they send no messages and Bob outputs a uniform bit. This protocol has error \(\varepsilon\) on every input and information cost \((1-2\varepsilon)m\) under independent uniform inputs. We refer to this as the mixture protocol.

Braverman, Garg, Pankratov, and Weinstein asked whether the information complexity, normalized by the input length \(m\), admits a lower bound \(1-\alpha\varepsilon-o(\varepsilon)\) for some constant \(\alpha\ge2\), for small constant error and sufficiently large dimension~\cite[Problem~1.4]{BGPW2016}. Their result shows that the normalized cost approaches one as the error decreases, without determining this first-order dependence. Dagan, Filmus, Hatami, and Li subsequently restated the question~\cite[Section~8]{DFHL2018}.

We resolve this question by showing that the mixture protocol is optimal up to a universal additive constant. Under independent uniform inputs, the zero-error information cost is \(m\) bits. Allowing error \(\varepsilon\) lets the mixture protocol save \(2\varepsilon m\) bits, while any protocol in which Bob produces the output and whose average error is at most \(\varepsilon\) has information cost at least \((1-2\varepsilon)m-c_0\), where \(c_0<4.772\). Thus, the optimal first-order coefficient is \(2\). Moreover, these bounds hold for every \(m\ge1\) and \(0\le\varepsilon\le1/2\), determining the information--error tradeoff to within a universal additive constant over the entire error range.

We also extend this characterization to matrix multiplication over finite fields. For independent uniform matrix inputs, we determine the information--error tradeoff up to an additive constant independent of the matrix dimensions, field size, and error. The information lower bounds for both inner product and matrix multiplication follow from a Shannon-to-min-entropy theorem for convolution on discrete groups. We show that the ordered product of sufficiently many independent random variables taking values in a discrete group and having finitely supported distributions has min-entropy at least the average Shannon entropy of their distributions minus a universal constant. The required number of factors is logarithmic in the largest support size. This holds even when the factors have different distributions and the group is infinite or noncommutative. Although high Shannon entropy alone does not rule out a large probability at one value, this inequality bounds the probability that the product equals any given element using only the factors' Shannon entropies.

\subsection{Main results}

\paragraph{Shannon-to-min-entropy conversion.}
For a discrete random variable \(X\) with distribution \(p\), write
\[
H(X)=\sum_{x:p(x)>0}p(x)\log_2\frac1{p(x)},
\qquad
H_\infty(X)=-\log_2\max_xp(x).
\]
All entropies are in bits, \(\ln\) denotes the natural logarithm, and \(a^+=\max\{a,0\}\). Let \(G\) be a discrete group, and let \(Z_1,\ldots,Z_k\) be independent random variables taking values in \(G\), with respective distributions \(Q_1,\ldots,Q_k\). The convolution \(Q_1*\cdots*Q_k\) is the distribution of their ordered product \(Z_1\cdots Z_k\). For a finitely supported distribution \(P\), its largest point probability is \(\|P\|_\infty=\max_g P(g)=2^{-H_\infty(P)}\).

\begin{theorem}[Shannon-to-min-entropy convolution bound]\label{thm:intro-flattening}
Let \(G\) be a discrete group, and let \(Q_1,\ldots,Q_k\) be finitely supported probability distributions on \(G\), where \(k\ge1\). Define
\[
s=\max_{1\le i\le k}|\supp Q_i|\ge2,
\qquad
n=\lfloor\log_2s\rfloor,
\qquad
\overline H=\frac1k\sum_{i=1}^kH(Q_i).
\]
Then
\begin{equation}\label{eq:intro-flattening}
\lVert Q_1*\cdots*Q_k\rVert_\infty
\le C_0\exp\!\bigl((n\ln2-k)^+\bigr)2^{-\overline H},
\qquad
C_0=16+8\sqrt2.
\end{equation}
Consequently, if \(k\ge\lceil n\ln2\rceil\), then
\[
H_\infty(Q_1*\cdots*Q_k)\ge\overline H-c_0,
\qquad
c_0=\log_2C_0<4.772.
\]
\end{theorem}

The loss \(c_0\) is independent of the group, the support sizes, and the distributions \(Q_i\), which need not be identical. The group can be infinite or noncommutative. We also prove a version for a Poisson number of factors drawn from a common distribution (Corollary~\ref{cor:poisson-flattening}), which is used to establish the information lower bound.

\paragraph{Information complexity of matrix multiplication.}
Let \(f:\mathcal X\times\mathcal Y\to\mathcal Z\) be a function to be computed, and let \(\widehat f\) denote the output of a finite randomized protocol. We consider two output conventions. Under the Bob-only convention, Bob can determine the output using his own input, the transcript \(\Pi\), and his private randomness. Under the transcript-determined convention, the output must be a function \(g(\Pi)\) of the transcript alone. For an input distribution \(\mu\), the average-error and pointwise-error constraints are, respectively,
\[
\Pr[\widehat f(\boldsymbol X,\boldsymbol Y)\ne f(\boldsymbol X,\boldsymbol Y)]\le\varepsilon,
\qquad
\max_{\boldsymbol x\in\mathcal X,\,\boldsymbol y\in\mathcal Y}
\Pr[\widehat f(\boldsymbol x,\boldsymbol y)\ne f(\boldsymbol x,\boldsymbol y)]\le\varepsilon.
\]
In the first probability, the inputs \((\boldsymbol X,\boldsymbol Y)\) are drawn from \(\mu\), whereas in the second, the input pair is fixed. Both probabilities include the randomness used by the protocol. For a finite set \(\mathcal S\), let \(U_{\mathcal S}\) denote the uniform distribution on \(\mathcal S\). For a product-uniform input distribution \(U_{\mathcal X\times\mathcal Y}\), write \(\IC_{U_{\mathcal X\times\mathcal Y}}^{\mathrm B,\mathrm{av}}(f,\varepsilon)\) and \(\IC_{U_{\mathcal X\times\mathcal Y}}^{\mathrm B,\mathrm{pw}}(f,\varepsilon)\) for the infimum of \eqref{eq:internal-information} over Bob-only protocols obeying the respective constraints. Under the product-uniform prior, the two error conditions coincide at zero error, and we denote their common complexity by \(\IC_{U_{\mathcal X\times\mathcal Y}}^{\mathrm B}(f,0)\).

Let \(q\) be a prime power, \(m\ge1\), and \(1\le t\le s\). Alice receives \(\boldsymbol X\in\F_q^{t\times m}\), Bob receives \(\boldsymbol Y\in\F_q^{s\times m}\), and Bob must output
\begin{equation}\label{eq:distributed-matrix-multiplication}
\MM_{t,s,m}(\boldsymbol X,\boldsymbol Y)=\boldsymbol X\boldsymbol Y^{\mathsf T}.
\end{equation}
For \(0\le\varepsilon\le1\), set
\begin{equation}\label{eq:alpha-matrix-multiplication}
\alpha_{q,t,s}(\varepsilon)=\left(1-\frac{q^{ts}}{q^{ts}-1}\varepsilon\right)^+.
\end{equation}
Let \(\boldsymbol\Lambda\) be uniform over the nonzero matrices in \(\F_q^{t\times s}\), and define
\begin{equation}\label{eq:kappa-matrix-multiplication}
\kappa_{q,t,s}=\E_{\boldsymbol\Lambda}\frac{t}{\rank\boldsymbol\Lambda}.
\end{equation}
We prove \(\kappa_{q,t,s}\le2\) in Lemma~\ref{lem:reciprocal-rank}.

\begin{theorem}[Information complexity of matrix multiplication]\label{thm:matrix-multiplication}
Let \(m\ge1\), \(1\le t\le s\), and \(U=U_{\F_q^{t\times m}\times\F_q^{s\times m}}\). For \(0\le\varepsilon\le1\),
\begin{equation}\label{eq:matrix-multiplication-bounds}
\left[\alpha_{q,t,s}(\varepsilon)mt\log_2q-c_0\kappa_{q,t,s}\right]^+
\le\IC_{U}^{\mathrm B,\mathrm{av}}(\MM_{t,s,m},\varepsilon)
\le\IC_{U}^{\mathrm B,\mathrm{pw}}(\MM_{t,s,m},\varepsilon)
\le\alpha_{q,t,s}(\varepsilon)mt\log_2q.
\end{equation}
The same lower bound holds when the transcript determines the output. At zero error,
\begin{equation}\label{eq:matrix-multiplication-zero-error}
\IC_{U}^{\mathrm B}(\MM_{t,s,m},0)=mt\log_2q.
\end{equation}
\end{theorem}
The upper and lower bounds differ by at most \(2c_0<9.544\) bits, independently of \(m,t,s,q\), and \(\varepsilon\).

\paragraph{Resolution of the binary inner-product problem.}
Taking \(q=2\) and \(t=s=1\) gives binary inner product. Every nonzero \(1\times1\) matrix has rank one, so \(\kappa_{2,1,1}=1\), preserving the smaller loss \(c_0\).

\begin{corollary}[Binary inner product]\label{cor:intro-binary}
Let \(m\ge1\) and \(U=U_{\F_2^m\times\F_2^m}\). For \(0\le\varepsilon\le1/2\),
\begin{equation}\label{eq:intro-binary-inner-product}
\bigl[(1-2\varepsilon)m-c_0\bigr]^+
\le\IC_{U}^{\mathrm B,\mathrm{av}}(\IP_m,\varepsilon)
\le\IC_{U}^{\mathrm B,\mathrm{pw}}(\IP_m,\varepsilon)
\le(1-2\varepsilon)m.
\end{equation}
\end{corollary}
The Bob-only lower bound also applies to the transcript-determined output convention used in~\cite{BGPW2016}. The mixture protocol satisfies this convention if Bob also sends his output bit to Alice, increasing its information cost by at most one bit.

The normalized form of \eqref{eq:intro-binary-inner-product} is
\begin{equation}\label{eq:binary-normalized-resolution}
\begin{aligned}
1-2\varepsilon-\frac{c_0}{m}
&\le\frac{\IC_{U}^{\mathrm B,\mathrm{av}}(\IP_m,\varepsilon)}{m}\\
&\le\frac{\IC_{U}^{\mathrm B,\mathrm{pw}}(\IP_m,\varepsilon)}{m}
\le1-2\varepsilon.
\end{aligned}
\end{equation}
Consequently, as \(m\to\infty\), both normalized information complexities converge to \(1-2\varepsilon\) uniformly over \(\varepsilon\in[0,1/2]\), with error at most \(c_0/m\). The exact zero-error cost of \(m\) bits follows from \eqref{eq:matrix-multiplication-zero-error} with \(q=2\) and \(t=s=1\). These bounds resolve the question of Braverman, Garg, Pankratov, and Weinstein~\cite[Problem~1.4]{BGPW2016} on the error dependence of binary inner-product information complexity, with the optimal first-order coefficient \(\alpha=2\).

\subsection{Related work}

Entropy inequalities for sums and differences on groups have been studied in additive combinatorics and information theory~\cite{Tao2010,MadimanKontoyiannis2018,MadimanWangWoo2021}. The min-entropy convolution inequalities of Xu, Melbourne, and Madiman~\cite{XuMelbourneMadiman2017} and the Rogozin-type inequalities of Madiman, Melbourne, and Xu~\cite{MadimanMelbourneXu2017} control output concentration from bounds on the input densities or largest point probabilities. The Rogozin-type results already allow independent, nonidentically distributed inputs. Theorem~\ref{thm:intro-flattening} instead starts from the inputs' average Shannon entropy and requires no input min-entropy bound. It directly bounds the probability that the ordered product of independent random variables on any discrete group equals any given element. This bound applies once the number of factors reaches a logarithmic threshold in the largest support size. Golubeva, Pan, and Tamuz~\cite{GolubevaPanTamuz2024} study the asymptotic rates \(\lim_{k\to\infty}H_\alpha(Q^{*k})/k\) for a fixed distribution \(Q\). Our bound concerns the output distribution at an explicit finite convolution count and allows the random variables to have different distributions.

Inner product is also a classical two-source extractor: Chor and Goldreich~\cite{ChorGoldreich1988} showed that binary inner product extracts an almost uniform bit from two independent sources with sufficient min-entropy. Strong extraction requires the output to remain close to uniform even when one source is revealed. Finite-field inner product satisfies such guarantees under min-entropy assumptions~\cite[Section~2]{DziembowskiKazanaObremski2013}. In particular, these guarantees already restrict predictors that depend on the revealed source \(\boldsymbol Y\). We instead bound the accuracy of predicting the inner product from one input in terms of both input distributions' Kullback--Leibler (KL) divergences from uniformity. Large Shannon entropy alone does not supply the min-entropy assumptions of the extractor bounds. The resulting inequality is linear in prediction error, up to a universal additive loss, so it can be averaged over the conditional input distributions arising from communication transcripts (Corollary~\ref{cor:decoder-aware}).

Braverman, Garg, Pankratov, and Weinstein developed information lower bounds through self-reducibility~\cite{BGPW2016}, and Dagan, Filmus, Hatami, and Li studied the tradeoff between information complexity and error~\cite{DFHL2018}. The binary inner-product bounds in Corollary~\ref{cor:intro-binary} resolve their coefficient question with a constant loss uniform in dimension and error. The posterior factorization and information-cost identity are standard in interactive information complexity~\cite{Braverman2015}, and the connection to amortized communication is due to Braverman and Rao~\cite{BravermanRao2014}.

Huang, Mao, Wang, Yang, and Zhang~\cite{HuangMaoWangYangZhang2025} use min-entropy to avoid square-root losses arising from Shannon-entropy and Pinsker-based analyses of multi-party communication problems. Their method develops a structure-versus-pseudorandomness decomposition to compensate for the absence of a min-entropy chain rule. Our use of min-entropy has a related motivation but a different role: the convolution bound controls a decoder's exponential moment, and the Gibbs variational formula then gives a prediction inequality in KL divergence. The resulting lower bound applies directly to information cost.

The entropy projection inequality follows from Shearer's inequality~\cite{ChungGrahamFranklShearer1986} by averaging over invertible changes of basis. Related fractional-cover formulations appear in Madiman and Tetali~\cite{MadimanTetali2010}. Its divergence form lets us combine scalar prediction bounds across output tests of different ranks while retaining a universal additive loss for matrix multiplication.

\subsection{Organization}

Section~\ref{sec:overview} outlines why accurate computation requires information about the inputs, first for binary inner product and then for matrix multiplication. After the preliminaries in Section~\ref{sec:preliminaries}, Section~\ref{sec:flattening} proves the convolution bounds needed for this argument, Section~\ref{sec:prediction} establishes the finite-field prediction inequality, and Section~\ref{sec:bilinear-ic} derives the information-complexity results. The appendices contain the threshold and sharpness estimates, the detailed Fourier calculation, and supporting protocol and rank arguments.

\section{Technical Overview}\label{sec:overview}

To prove the lower bound on the information complexity of binary inner product, we examine what Bob can predict after a transcript has been fixed. When the inputs are independent and uniform, the protocol's information cost equals the average, over transcripts, of the sum of the Kullback--Leibler divergences of the conditional input distributions from their original uniform distributions. These divergences quantify how far the input distributions depart from uniformity after the transcript is observed. At a fixed transcript, Bob's output can be treated as a prediction based on his own input, so the lower bound reduces to an inequality relating the prediction error to these divergences. We establish this inequality using a convolution bound that controls the probability that a sum of independent random variables takes any particular value. We then extend the argument to finite fields and matrix multiplication. For matrix multiplication, the entries of the output share the same inputs, so information lower bounds for individual entries cannot simply be added. Finally, we explain why the convolution bound incurs only a constant loss of entropy relative to Shannon entropy.

\subsection{From protocol posteriors to prediction}

We first consider binary inner product with independent uniform inputs \(\boldsymbol X,\boldsymbol Y\in\F_2^m\). Let \(\Pi\) denote the transcript of a protocol for this problem. Their joint input distribution is \(U_{\F_2^m\times\F_2^m}\). For any transcript value \(\pi\) of positive probability, the inputs remain independent after conditioning on \(\Pi=\pi\) (Lemma~\ref{lem:posterior-product}). Their conditional distributions, called the posteriors, need not remain uniform. Moreover, by Lemma~\ref{lem:posterior-product}, the information cost of the protocol satisfies
\[
\IC_{U_{\F_2^m\times\F_2^m}}(\Pi)=\E_\Pi\bigl[D(P_{\boldsymbol X\mid\Pi}\Vert U_{\F_2^m})+D(P_{\boldsymbol Y\mid\Pi}\Vert U_{\F_2^m})\bigr],
\]
where \(U_{\F_2^m}\) denotes the uniform distribution on \(\F_2^m\) and \(D\) denotes Kullback--Leibler divergence. For any distribution \(P\) on \(\F_2^m\), we have \(D(P\Vert U_{\F_2^m})=m-H(P)\).
\par

We apply this identity to a Bob-only protocol for binary inner product with average error at most \(\varepsilon\) under \(U_{\F_2^m\times\F_2^m}\). By Lemma~\ref{lem:deterministic-decoding}, we can replace Bob's output by a deterministic decoder \(z(\boldsymbol Y,\Pi)\) without increasing the average error or changing the information cost. Once a transcript \(\pi\) is fixed, the decoder becomes a predictor \(z_\pi(\boldsymbol Y)=z(\boldsymbol Y,\pi)\) that depends only on Bob's input. Its conditional error is
\[
\eta_\pi=\Pr[z_\pi(\boldsymbol Y)\ne\boldsymbol X^{\mathsf T}\boldsymbol Y\mid\Pi=\pi].
\]
Since the protocol has average error at most \(\varepsilon\), we have \(\E_\Pi\eta_\Pi\le\varepsilon\).

We can therefore obtain the information lower bound from an inequality that no longer involves a communication protocol. Let \(\boldsymbol X,\boldsymbol Y\in\F_2^m\) be any independent random vectors, not necessarily uniform, and let \(z(\boldsymbol Y)\) be any predictor of their inner product. Writing \(\eta=\Pr[z(\boldsymbol Y)\ne\boldsymbol X^{\mathsf T}\boldsymbol Y]\), we will prove in Corollary~\ref{cor:decoder-aware} that

\begin{equation}\label{eq:overview-binary-prediction}
D(P_{\boldsymbol X}\Vert U_{\F_2^m})+D(P_{\boldsymbol Y}\Vert U_{\F_2^m})\ge m(1-2\eta)-c_0.
\end{equation}
For each transcript \(\pi\), we can apply this inequality to the posterior distributions and the predictor \(z_\pi\). Averaging over transcripts yields
\[
\begin{aligned}
\IC_{U_{\F_2^m\times\F_2^m}}(\Pi)
&=\E_\Pi\bigl[D(P_{\boldsymbol X\mid\Pi}\Vert U_{\F_2^m})+D(P_{\boldsymbol Y\mid\Pi}\Vert U_{\F_2^m})\bigr]\\
&\ge\E_\Pi[m(1-2\eta_\Pi)-c_0]\\
&\ge m(1-2\varepsilon)-c_0.
\end{aligned}
\]
\par

It remains to establish \eqref{eq:overview-binary-prediction}, which we do in the next subsection using the convolution bound.

\subsection{From prediction to a Poisson point probability}\label{sec:prediction-overview}

We now prove \eqref{eq:overview-binary-prediction}, starting with an optimal predictor. For each \(\boldsymbol y\in\F_2^m\), let \(\phi(\boldsymbol y)\) be a most likely value of \(\boldsymbol X^{\mathsf T}\boldsymbol y\). Since \(\boldsymbol X\) and \(\boldsymbol Y\) are independent, this choice minimizes the overall prediction error. For each fixed \(\boldsymbol x\), define
\[
f(\boldsymbol x)=1-2\Pr[\boldsymbol x^{\mathsf T}\boldsymbol Y\ne\phi(\boldsymbol Y)].
\]
Thus, \(f(\boldsymbol x)\) is the difference between the probabilities of a correct and an incorrect prediction when the first input is fixed to \(\boldsymbol x\). Let \(\eta_*=\Pr[\boldsymbol X^{\mathsf T}\boldsymbol Y\ne\phi(\boldsymbol Y)]\) be the error probability of the optimal predictor \(\phi\). Then \(\E_{\boldsymbol X}f(\boldsymbol X)=1-2\eta_*\). Applying the Gibbs variational formula~\cite[Lemma~4.10]{vanHandel2016} gives
\[
m(1-2\eta_*)\le D(P_{\boldsymbol X}\Vert U_{\F_2^m})+\log_2\E_{\boldsymbol U\sim U_{\F_2^m}}2^{mf(\boldsymbol U)}.
\]
Therefore, to obtain \eqref{eq:overview-binary-prediction}, it suffices to bound the remaining exponential moment by
\[
\log_2\E_{\boldsymbol U\sim U_{\F_2^m}}2^{mf(\boldsymbol U)}
\le D(P_{\boldsymbol Y}\Vert U_{\F_2^m})+c_0
=m-H(\boldsymbol Y)+c_0.
\]

The key is to express this exponential moment in terms of the distribution of a random sum, whose largest point probability can be bounded using our convolution bound. First, encode \(\boldsymbol Y\) together with its predicted value as \(\boldsymbol Z=(\boldsymbol Y,\phi(\boldsymbol Y))\in\F_2^{m+1}\). For any fixed \(\boldsymbol x\), the expression \(\boldsymbol x^{\mathsf T}\boldsymbol Y+\phi(\boldsymbol Y)\) equals zero precisely when the prediction is correct. Consequently, the sign \((-1)^{(\boldsymbol x,1)^{\mathsf T}\boldsymbol Z}\) is \(+1\) for a correct prediction and \(-1\) otherwise, giving
\[
\E_{\boldsymbol Z}(-1)^{(\boldsymbol x,1)^{\mathsf T}\boldsymbol Z}
=\Pr[\boldsymbol x^{\mathsf T}\boldsymbol Y=\phi(\boldsymbol Y)]
-\Pr[\boldsymbol x^{\mathsf T}\boldsymbol Y\ne\phi(\boldsymbol Y)]
=f(\boldsymbol x).
\]

We need to bound an exponential in \(f(\boldsymbol x)\), whereas the preceding identity gives an expectation equal to \(f(\boldsymbol x)\). The Poisson generating function allows us to obtain the required exponential from independent copies of \(\boldsymbol Z\). Indeed, for independent copies \(\boldsymbol Z_1,\boldsymbol Z_2,\ldots\), we have
\[
\begin{aligned}
\E(-1)^{(\boldsymbol x,1)^{\mathsf T}\sum_{i=1}^k\boldsymbol Z_i}
&=\prod_{i=1}^k\E(-1)^{(\boldsymbol x,1)^{\mathsf T}\boldsymbol Z_i}\\
&=f(\boldsymbol x)^k.
\end{aligned}
\]
If the number of copies is Poisson distributed, averaging these powers produces an exponential. Specifically, let \(K\) be a Poisson random variable with mean \(m\ln2\), independent of the copies, and define \(\boldsymbol S=\sum_{i=1}^K\boldsymbol Z_i=(\boldsymbol S',S_0)\), where \(\boldsymbol S'\in\F_2^m\) and \(S_0\in\F_2\). The empty sum is zero. Conditioning on \(K\) and using the Poisson generating function \eqref{eq:poisson-pgf} in Appendix~\ref{sec:standard-tools} gives
\par

\[
\E_{\boldsymbol S}(-1)^{\boldsymbol x^{\mathsf T}\boldsymbol S'+S_0}
=\E_K f(\boldsymbol x)^K
=e^{m\ln2(f(\boldsymbol x)-1)}
=2^{m(f(\boldsymbol x)-1)}.
\]
Thus, the chosen Poisson mean produces exactly the exponential needed in the preceding bound, up to a factor of \(2^{-m}\).

Let \(\boldsymbol U\) be uniform on \(\F_2^m\) and independent of \(\boldsymbol S\). For any fixed \(\boldsymbol s\in\F_2^m\), the average of \((-1)^{\boldsymbol U^{\mathsf T}\boldsymbol s}\) is one when \(\boldsymbol s=\boldsymbol0\) and zero otherwise, since the positive and negative terms cancel whenever \(\boldsymbol s\ne\boldsymbol0\). Hence, only the outcomes with \(\boldsymbol S'=\boldsymbol0\) contribute. Averaging the preceding identity over \(\boldsymbol U\), we obtain
\begin{align}\label{eq:overview-binary-moment}
2^{-m}\E_{\boldsymbol U}2^{mf(\boldsymbol U)}
&=\E_{\boldsymbol S}\bigl[\1_{\{\boldsymbol S'=\boldsymbol0\}}(-1)^{S_0}\bigr]\notag\\
&=\Pr[\boldsymbol S=(\boldsymbol0,0)]-\Pr[\boldsymbol S=(\boldsymbol0,1)]\notag\\
&\le\|P_{\boldsymbol S}\|_\infty.
\end{align}
We have therefore reduced the exponential moment bound to controlling the largest point probability of \(\boldsymbol S\).

This is where the convolution bound applies. Since \(\boldsymbol Z\) contains \(\boldsymbol Y\) and its last coordinate is determined by \(\boldsymbol Y\), we have \(H(\boldsymbol Z)=H(\boldsymbol Y)\) and \(|\supp P_{\boldsymbol Z}|\le2^m\). The Poisson mean \(m\ln2\) is therefore large enough to apply Corollary~\ref{cor:poisson-flattening}, which gives
\[
\|P_{\boldsymbol S}\|_\infty\le C_0 2^{-H(\boldsymbol Z)}=C_0 2^{-H(\boldsymbol Y)}
\]
when the support of \(\boldsymbol Z\) has at least two points. If \(\boldsymbol Z\) is deterministic, then \(H(\boldsymbol Y)=0\), and the same bound follows from \(\|P_{\boldsymbol S}\|_\infty\le1\). Combining this with \eqref{eq:overview-binary-moment} and taking logarithms yields
\[
\log_2\E_{\boldsymbol U}2^{mf(\boldsymbol U)}\le m-H(\boldsymbol Y)+c_0.
\]
Substituting this bound into the Gibbs inequality proves \eqref{eq:overview-binary-prediction} for the optimal predictor. Since any specified decoder has error \(\eta\ge\eta_*\), the inequality also holds for its error.

The prediction inequality extends to arbitrary finite fields. For independent random vectors \(\boldsymbol X,\boldsymbol Y\in\F_q^m\) and any predictor \(z(\boldsymbol Y)\) with error probability \(\eta\), Corollary~\ref{cor:decoder-aware} gives
\[
D(P_{\boldsymbol X}\Vert U_{\F_q^m})
+D(P_{\boldsymbol Y}\Vert U_{\F_q^m})
\ge m\log_2q\left(1-\frac q{q-1}\eta\right)-c_0.
\]
The proof uses an averaging identity over \(\F_q\) in place of the binary sign calculation. We give the full argument in Section~\ref{sec:prediction}, with the Fourier calculation in Appendix~\ref{sec:prediction-fourier-details}.
\par

\subsection{Testing a matrix output without accumulating losses}

The prediction inequality from the previous subsection concerns a single inner product, whereas matrix multiplication requires Bob to output the entire matrix \(\boldsymbol X\boldsymbol Y^{\mathsf T}\), where \(\boldsymbol X\in\F_q^{t\times m}\) and \(\boldsymbol Y\in\F_q^{s\times m}\). Since all entries of this matrix depend on the same inputs, we cannot obtain a lower bound for the whole matrix by simply adding the scalar information lower bounds for its entries.
Instead, we reduce the matrix error to a scalar prediction error using a random linear test.
For matrices \(\boldsymbol A,\boldsymbol B\in\F_q^{t\times s}\), write
\[
\langle\boldsymbol A,\boldsymbol B\rangle_{\mathrm F}
=\sum_{i=1}^t\sum_{j=1}^s A_{ij}B_{ij}.
\]
Choose a uniform nonzero matrix \(\boldsymbol\Lambda\in\F_q^{t\times s}\), independently of the inputs and transcript, and consider the scalar \(\langle\boldsymbol\Lambda,\boldsymbol X\boldsymbol Y^{\mathsf T}\rangle_{\mathrm F}\).
\par

 For any nonzero error matrix \(\boldsymbol E\), the map \(\boldsymbol\Lambda\mapsto\langle\boldsymbol\Lambda,\boldsymbol E\rangle_{\mathrm F}\) is a nonzero linear functional. Consequently, exactly \(q^{ts-1}(q-1)\) of the \(q^{ts}-1\) nonzero tests detect the error. At a fixed transcript \(\pi\), let \(\eta\) be the probability that Bob's matrix output is incorrect, and let \(\eta_{\boldsymbol\Lambda}\) be the probability that applying the test to his output gives an incorrect scalar value. Averaging over the test matrices yields
\[
\E_{\boldsymbol\Lambda}\eta_{\boldsymbol\Lambda}
=\eta\frac{q^{ts-1}(q-1)}{q^{ts}-1}.
\]

To apply the scalar prediction inequality, we express the tested quantity as an inner product of two projected inputs. Suppose \(\boldsymbol\Lambda\) has rank \(d\). Choose uniformly a factorization \(\boldsymbol\Lambda=\boldsymbol L^{\mathsf T}\boldsymbol R\), where \(\boldsymbol L\in\F_q^{d\times t}\) and \(\boldsymbol R\in\F_q^{d\times s}\) both have full row rank. Then
\[
\langle\boldsymbol\Lambda,\boldsymbol X\boldsymbol Y^{\mathsf T}\rangle_{\mathrm F}
=\langle\boldsymbol L\boldsymbol X,\boldsymbol R\boldsymbol Y\rangle_{\mathrm F}.
\]
Thus, the scalar test is an inner product between two matrices that can be viewed as vectors in \(\F_q^{dm}\). For fixed \(\boldsymbol L\) and \(\boldsymbol R\), these projected inputs remain independent under the posterior distributions. Although Bob's tested output may depend on his entire input \(\boldsymbol Y\), the target depends on it only through \(\boldsymbol R\boldsymbol Y\). Conditional on this projection, the remaining randomness in \(\boldsymbol Y\) is independent of \(\boldsymbol L\boldsymbol X\), so Bob's output can be viewed as a randomized prediction based only on \(\boldsymbol R\boldsymbol Y\). Replacing it by a most likely value of the target cannot increase the error. We can therefore apply the scalar prediction inequality to the projected inputs.

It remains to relate the divergences of these projected inputs to those of the original inputs. Write \(\Delta_{\boldsymbol X}\) and \(\Delta_{\boldsymbol Y}\) for the divergences of the two posterior distributions from their respective uniform priors. Conditional on rank \(d\), the factors \(\boldsymbol L\) and \(\boldsymbol R\) are independent and uniform over full-row-rank matrices of their respective sizes, because every rank-\(d\) matrix admits the same number of such factorizations. We can therefore use an entropy inequality for random linear projections. Shearer's inequality states that the average entropy of a uniformly chosen subset of \(d\) rows is at least a \(d/t\) fraction of the entropy of all \(t\) rows. Applying this inequality after a random invertible change of basis shows that the expected divergence of \(\boldsymbol L\boldsymbol X\) from uniformity is at most \(d\Delta_{\boldsymbol X}/t\). Similarly, the expected divergence of \(\boldsymbol R\boldsymbol Y\) is at most \(d\Delta_{\boldsymbol Y}/s\). On the other hand, the scalar prediction inequality gives a lower bound on the sum of these projected divergences in terms of the prediction error. Combining the two bounds and dividing by \(d\) yields
\[
\frac{\Delta_{\boldsymbol X}}t+\frac{\Delta_{\boldsymbol Y}}s
\ge m\log_2q\left(1-\frac q{q-1}\E[\eta_{\boldsymbol\Lambda}\mid\rank\boldsymbol\Lambda=d]\right)-\frac{c_0}d.
\]

We now average this inequality over the rank of \(\boldsymbol\Lambda\). By the first identity, the average prediction error satisfies
\[
\frac q{q-1}\E_{\boldsymbol\Lambda}\eta_{\boldsymbol\Lambda}
=\frac{q^{ts}}{q^{ts}-1}\eta.
\]
The additive loss becomes \(c_0\E[1/\rank\boldsymbol\Lambda]\). Since \(t\le s\) and the divergences are nonnegative, multiplying the resulting inequality by \(t\) gives
\[
\Delta_{\boldsymbol X}+\Delta_{\boldsymbol Y}
\ge mt\log_2q\left(1-\frac{q^{ts}}{q^{ts}-1}\eta\right)
-c_0\kappa_{q,t,s},
\]
where \(\kappa_{q,t,s}=\E[t/\rank\boldsymbol\Lambda]\le2\). Thus, the additive loss is bounded by \(2c_0\), independently of the matrix dimensions. Finally, averaging over transcripts converts the two posterior divergences into the protocol's information cost and the conditional matrix error into its average error, proving the desired information lower bound. At zero error, the exact scalar prediction bound removes the additive loss. Section~\ref{sec:bilinear-ic} gives the full argument, including the projection inequality and the exact zero-error case.

\subsection{\texorpdfstring{Entropy loss in the convolution bound}{Entropy loss in the convolution bound}}

The convolution bound in Theorem~\ref{thm:intro-flattening} and its Poisson counterpart, stated later in Corollary~\ref{cor:poisson-flattening}, control the largest point probability of a product using the Shannon entropy of its factors. The difficulty is that Shannon entropy alone does not control the largest point probability of a single distribution: a distribution can have high entropy while assigning substantial probability to one point. To overcome this difficulty, we express each distribution as a mixture of uniform distributions. Specifically, we order the support points of a distribution \(Q\) by decreasing probability and represent \(Q\) as a mixture of uniform distributions on nested sets. Let \(J\) be the size of the randomly selected set, and define \(R=\lfloor\log_2J\rfloor\). The entropy of \(Q\) satisfies
\[
H(Q)\le H(R)+\E R+1.
\]
Now consider a product of independent random variables, with one uniform component selected for each factor. Its largest point probability is at most the reciprocal of the largest selected set size, since convolution cannot increase the largest point probability of any factor. Thus, writing \(M\) for the maximum of the selected scales, the convolution is bounded by \(\E2^{-M}\). To obtain a bound in terms of \(H(Q)\), we need to control this expectation using both \(\E R\) and \(H(R)\). A bound that retains only \(\E R\) would discard the entropy contributed by the random choice of scale.

We first explain the argument for a Poisson number of factors, as required by the prediction inequality. Let \(n=\lfloor\log_2|\supp Q|\rfloor\ge1\), and choose the number of factors according to a Poisson distribution with mean \(n\ln2\). Each factor has an independently selected scale distributed as \(R\), and \(M\) is their maximum, with \(M=0\) when no factor is selected. For \(0\le r<n\), the Poisson generating function gives
\[
\Pr[M\le r]=2^{-n\Pr[R\ge r+1]}.
\]
Since \(0\le M\le n\), expressing \(2^{-M}\) in terms of the events \(M\le r\) yields
\[
\E2^{-M}=2^{-n}+\frac12\sum_{r=0}^{n-1}2^{-r-n\Pr[R\ge r+1]}.
\]
The remaining task is to bound this sum while retaining the contribution of \(H(R)\).

The key observation is that the exponents in this sum can be expressed through distances on a cycle. Identify the endpoints \(0\) and \(n\), and let \(\overline R=R\bmod n\). For each \(0\le r<n\), let \(d_r\) be the expected clockwise distance from \(\overline R\) to \(r\). When \(R\le r\), this distance is \(r-R\); otherwise, it is \(r-R+n\). Taking expectations gives
\[
r+n\Pr[R\ge r+1]=\E R+d_r.
\]
Consequently,
\[
\E2^{-M}=2^{-n}+\frac12\cdot2^{-\E R}\sum_{r=0}^{n-1}2^{-d_r}.
\]
This separates the contribution of the average scale from the remaining sum. The directed-cycle estimate bounds \(\sum_r2^{-d_r}\) by a universal constant times \(2^{-H(\overline R)}\). Its proof compares the distribution of a random clockwise distance with a geometric distribution. Since identifying the endpoints merges only the values \(0\) and \(n\), we have \(H(\overline R)\ge H(R)-1\). A separate entropy estimate, \(H(R)+\E R\le n+1\), controls the remaining term \(2^{-n}\). Combining these estimates gives
\[
\E2^{-M}\le(8+4\sqrt2)2^{-H(R)-\E R}
\le C_0 2^{-H(Q)},
\]
where the last inequality follows from the entropy bound for the selected scale. This proves the desired Poisson convolution bound with a universal constant loss.

The argument also applies to a fixed number of independent factors whose distributions need not be identical. In this case, let \(R_i\) be the selected scale of the \(i\)-th factor, and let \(R_*\) have the average of their distributions. Independence gives
\[
\Pr[M\le r]
=\prod_i\Pr[R_i\le r]
\le e^{-k\Pr[R_*\ge r+1]}.
\]
When \(k\ge n\ln2\), this leads to the same threshold sum with \(R_*\) in place of \(R\).
By concavity of entropy and the entropy bound for each layer scale,
\[
\begin{aligned}
H(R_*)+\E R_*
&\ge\frac1k\sum_{i=1}^k\bigl(H(R_i)+\E R_i\bigr)\\
&\ge\frac1k\sum_{i=1}^k H(Q_i)-1.
\end{aligned}
\]
Thus, the bound retains the average Shannon entropy of the input distributions.
\par

 This completes the argument underlying Theorem~\ref{thm:intro-flattening}.

\section{Preliminaries}\label{sec:preliminaries}

We collect the notation used in the entropy and convolution arguments. All entropies and KL divergences are measured in bits. We write \(L_q=\log_2q\), use \(\ln\) for the natural logarithm, and set \(a^+=\max\{a,0\}\). For a random object \(X\), let \(P_X\) denote its distribution, and write \(H(P_X)=H(X)\) for its Shannon entropy. We use \(U_{\mathcal X}\) for the uniform distribution on a finite set \(\mathcal X\). For independent uniform inputs on \(\mathcal X\) and \(\mathcal Y\), the joint distribution is \(U_{\mathcal X\times\mathcal Y}=U_{\mathcal X}\times U_{\mathcal Y}\). Random vectors and their realizations are denoted by bold uppercase and lowercase letters, respectively. Matrices are denoted by bold uppercase letters, with randomness specified when relevant.

For distributions \(P,Q\) on the same countable set, their KL divergence is
\[
D(P\Vert Q)=\sum_xP(x)\log_2\frac{P(x)}{Q(x)},
\]
where terms with \(P(x)=0\) are omitted, and \(D(P\Vert Q)=+\infty\) if \(P(x)>0\) and \(Q(x)=0\) for some \(x\). When \(P\) is a distribution on a finite set \(\mathcal X\), its divergence from the uniform distribution satisfies
\begin{equation}\label{eq:uniform-divergence}
D(P\Vert U_{\mathcal X})=\log_2|\mathcal X|-H(P).
\end{equation}

For the convolution results, let \(G\) be a discrete group, with its operation written multiplicatively and identity element \(e\). Given probability distributions \(\mu\) and \(\nu\) on \(G\), their convolution is defined by
\[
(\mu*\nu)(g)=\sum_{h\in G}\mu(h)\nu(h^{-1}g),
\qquad g\in G.
\]
Equivalently, if \(X\) and \(Y\) are independent random variables with distributions \(\mu\) and \(\nu\), respectively, then \(\mu*\nu\) is the distribution of their product \(XY\). The order of multiplication can matter when \(G\) is noncommutative. We write \(\|\mu\|_\infty=\sup_{g\in G}\mu(g)\) for the largest point probability of \(\mu\).

Now let \(Q\) be a probability distribution on \(G\), and let \(X_1,X_2,\ldots\) be independent random variables with common distribution \(Q\). For \(k\ge1\), the ordered product \(X_1\cdots X_k\) has distribution
\[
Q^{*k}=\underbrace{Q*\cdots*Q}_{k\text{ factors}}.
\]
We define the empty product to be \(e\), so that \(Q^{*0}=\delta_e\), where \(\delta_e\) assigns probability one to \(e\).

We also consider products with a random number of factors. For \(\lambda\ge0\), let \(K_\lambda\) have the Poisson distribution with mean \(\lambda\), independently of \(X_1,X_2,\ldots\). Let \(P_\lambda\) denote the distribution of the random product \(X_1\cdots X_{K_\lambda}\). Conditioning on \(K_\lambda\) gives
\[
P_\lambda=\sum_{k=0}^{\infty}\Pr[K_\lambda=k]Q^{*k}
=e^{-\lambda}\sum_{k=0}^{\infty}\frac{\lambda^k}{k!}Q^{*k}.
\]
This is the compound-Poisson distribution associated with \(Q\) and \(\lambda\). It is a probability-weighted mixture of convolution powers, including \(Q^{*0}\) when no factor is selected. Standard Poisson and finite-field Fourier identities are recorded in Appendix~\ref{sec:standard-tools}.

\section{Shannon-to-Min-Entropy Conversion on Discrete Groups}\label{sec:flattening}

We prove Theorem~\ref{thm:intro-flattening} by expressing each input distribution as a mixture of uniform distributions on nested sets. For a fixed choice of uniform components, the largest point probability of their convolution is controlled by the largest support size among them. The main task is to show that this bound remains effective after averaging over the choices of components. This requires accounting for both the sizes of the selected sets and the randomness in their selection. We first establish the corresponding entropy estimate, then use it to prove the convolution bound for a fixed number of factors and its compound-Poisson counterpart.

\subsection{Uniform layers and the maximum scale}

Let \(Q\) be a finitely supported probability distribution on a discrete group \(G\). List its support as \(g_1,\ldots,g_\ell\) in nonincreasing order of probability, so that
\[
q_1\ge\cdots\ge q_\ell>0,
\qquad
q_j=Q(g_j),
\]
and set \(q_{\ell+1}=0\). For \(1\le j\le\ell\), let \(\mathcal A_j=\{g_1,\ldots,g_j\}\). These sets are nested, with \(|\mathcal A_j|=j\). We call the uniform distributions \(U_{\mathcal A_j}\) on these sets the uniform layers of \(Q\). Define
\[
w_j=j(q_j-q_{j+1}),
\qquad 1\le j\le \ell.
\]

\begin{lemma}[Layer decomposition]\label{lem:layer-decomposition}
The numbers \(w_1,\ldots,w_\ell\) form a probability distribution, and
\begin{equation}\label{eq:layer-decomposition}
Q=\sum_{j=1}^\ell w_jU_{\mathcal A_j}.
\end{equation}
\end{lemma}

\begin{proof}
Since the probabilities \(q_j\) are nonincreasing, the weights \(w_j\) are nonnegative. Summation by parts gives
\[
\sum_{j=1}^\ell w_j
=\sum_{j=1}^\ell j(q_j-q_{j+1})
=\sum_{j=1}^\ell q_j=1.
\]
Moreover, \(g_i\) belongs to \(\mathcal A_j\) exactly when \(j\ge i\). Hence, the mixture on the right-hand side of \eqref{eq:layer-decomposition} assigns probability
\[
\sum_{j=i}^\ell\frac{w_j}{j}
=\sum_{j=i}^\ell(q_j-q_{j+1})
=q_i
\]
to \(g_i\). Thus, the mixture equals \(Q\).
\end{proof}

The decomposition gives a way to sample a random variable \(Y\) with distribution \(Q\): first choose a layer size \(J\) according to \(\Pr[J=j]=w_j\), then draw \(Y\) uniformly from \(\mathcal A_J\). We use the scale
\[
R=\lfloor\log_2J\rfloor
\]
to record the order of magnitude of the selected layer size. The next lemma relates the entropy of \(Q\) to the distribution of \(R\).

\begin{lemma}[Entropy of the layer scale]\label{lem:layer-entropy}
The layer construction satisfies
\begin{equation}\label{eq:layer-entropy}
H(Q)\le H(R)+\E R+1.
\end{equation}
\end{lemma}

\begin{proof}
Conditional on \(R=r\), the selected layer has size \(J<2^{r+1}\). Since the sets \(\mathcal A_j\) are nested, \(Y\) is then supported on fewer than \(2^{r+1}\) points. Therefore,
\[
H(Y\mid R=r)\le r+1.
\]
Averaging over \(R\) gives \(H(Y\mid R)\le\E R+1\). Since \(Y\) has distribution \(Q\), the entropy chain rule yields
\[
H(Q)=H(Y)\le H(R)+H(Y\mid R)
\le H(R)+\E R+1.
\]
\end{proof}

We now apply the layer decomposition to independent random variables with distributions \(Q_1,\ldots,Q_k\). For each \(Q_i\), fix an ordering of its support by decreasing probability, breaking ties arbitrarily, and let \(\mathcal A_{i,j}\) consist of its first \(j\) points. Conditional on independently selected layer sizes \(j_1,\ldots,j_k\), the factors are uniformly distributed on \(\mathcal A_{1,j_1},\ldots,\mathcal A_{k,j_k}\), respectively.

Convolution with a probability distribution cannot increase the largest point probability. Since \(\|U_{\mathcal A_{i,j_i}}\|_\infty=1/j_i\), Young's convolution inequality~\cite[Eq.~(1.1)]{CharalambidesChrist2011} gives
\begin{equation}\label{eq:conditional-layer-bound}
\left\lVert U_{\mathcal A_{1,j_1}}*\cdots*U_{\mathcal A_{k,j_k}}\right\rVert_\infty
\le\frac1{\max_i j_i}
\le2^{-\max_i r_i},
\end{equation}
where \(r_i=\lfloor\log_2j_i\rfloor\). Thus, the largest selected layer controls the point probabilities of the conditional convolution.

Set \(s=\max_i|\supp Q_i|\ge2\) and \(n=\lfloor\log_2s\rfloor\). Let \(R_1,\ldots,R_k\) be the independently selected scales, and define \(M=\max_iR_i\). By \eqref{eq:conditional-layer-bound}, averaging over the selected layers bounds the convolution norm by \(\E2^{-M}\). Since \(M\) takes values in \(\{0,\ldots,n\}\), summing the successive differences of \(2^{-r}\) and taking expectations gives
\begin{equation}\label{eq:maximum-threshold-identity}
\E2^{-M}=2^{-n}+\frac12\sum_{r=0}^{n-1}2^{-r}\Pr[M\le r].
\end{equation}
We have therefore reduced the convolution bound to estimating the probabilities that all selected scales are at most \(r\).

\subsection{An entropy bound for the threshold sum}

The threshold identity \eqref{eq:maximum-threshold-identity} allows us to exploit the independence of the selected scales. As we will see in the proof of Theorem~\ref{thm:intro-flattening}, the probability \(\Pr[M\le r]\) can be bounded by an exponential in the average tail probability of the scales. The factor \(n\ln2\) is useful because \(e^{-n\ln2\,p}=2^{-np}\), which matches the powers of two in \eqref{eq:maximum-threshold-identity}. This leads to the following estimate, which retains the entropy of the scale distribution.

\begin{lemma}[Threshold-sum bound]\label{lem:threshold-sum}
Let \(R\) be supported on \(\{0,\ldots,n\}\), where \(n\ge1\). Then
\begin{equation}\label{eq:threshold-sum}
2^{-n}+\frac12\sum_{r=0}^{n-1}2^{-r-n\Pr[R\ge r+1]}
\le(8+4\sqrt2)2^{-H(R)-\E R}.
\end{equation}
\end{lemma}

\paragraph{Proof sketch.}

The key step is to express the exponents in the sum as expected distances on a directed cycle. Identify the endpoints \(0\) and \(n\), and let \(\overline R=R\bmod n\) be the resulting random point on the cycle \(\{0,\ldots,n-1\}\). For \(0\le r<n\), the clockwise distance from \(\overline R\) to \(r\) is \(r-R\) when \(R\le r\), and \(r-R+n\) otherwise. Its expectation is therefore
\[
d_r=r-\E R+n\Pr[R\ge r+1].
\]
Consequently,
\[
\frac12\sum_{r=0}^{n-1}2^{-r-n\Pr[R\ge r+1]}
=\frac12\cdot2^{-\E R}\sum_{r=0}^{n-1}2^{-d_r}.
\]

Lemma~\ref{lem:cycle} bounds the sum on the right in terms of \(H(\overline R)\). Its proof compares the distribution of a random clockwise distance with a geometric distribution and gives
\[
\sum_{r=0}^{n-1}2^{-d_r}
\le(6+4\sqrt2)2^{-H(\overline R)}.
\]
The map \(R\mapsto\overline R\) preserves the value of \(R\) except that it identifies \(0\) and \(n\). Hence, \(H(R\mid\overline R)\le1\), so \(H(\overline R)\ge H(R)-1\). Combining these inequalities yields
\[
\frac12\sum_{r=0}^{n-1}2^{-r-n\Pr[R\ge r+1]}
\le(6+4\sqrt2)2^{-H(R)-\E R}.
\]

It remains to bound the term \(2^{-n}\). The endpoint estimate \eqref{eq:threshold-endpoint}, proved by comparing \(n-R\) with a geometric distribution, gives \(H(R)+\E R\le n+1\), and hence
\[
2^{-n}\le2\cdot2^{-H(R)-\E R}.
\]
Adding the two bounds proves \eqref{eq:threshold-sum}. Appendix~\ref{sec:threshold-details} contains the complete proof of the directed-cycle estimate and the endpoint bound.

\subsection{Proof of Theorem~\ref{thm:intro-flattening}}

\begin{proof}
Let \(Q_1,\ldots,Q_k\) be as in the theorem, with \(s,n\) and \(\overline H=k^{-1}\sum_iH(Q_i)\) defined there. For each \(i\), apply Lemma~\ref{lem:layer-decomposition} and independently select a uniform layer of \(Q_i\) according to its mixture weights. Let \(J_i\) be the size of the selected layer, and set \(R_i=\lfloor\log_2J_i\rfloor\) and \(M=\max_iR_i\). Since every selected layer has size at most \(s\), we have \(0\le M\le n\).

Conditional on the selected layers, \eqref{eq:conditional-layer-bound} bounds the largest point probability of the product by \(2^{-M}\). Averaging over the layer choices and taking the supremum over group elements gives
\begin{equation}\label{eq:layer-average-bound}
\lVert Q_1*\cdots*Q_k\rVert_\infty\le\E2^{-M}.
\end{equation}
We will bound this expectation using Lemma~\ref{lem:threshold-sum} and then relate the resulting entropy terms to \(\overline H\).

The scales \(R_i\) need not have identical distributions. To express their combined tail probabilities through a single distribution, let \(R_*\) have the average of their laws:
\[
\Pr[R_*=r]=\frac1k\sum_{i=1}^k\Pr[R_i=r],
\qquad 0\le r\le n.
\]
Since the scales are independent, for \(0\le r<n\) we have
\begin{align*}
\Pr[M\le r]
&=\prod_{i=1}^k\bigl(1-\Pr[R_i\ge r+1]\bigr)\\
&\le e^{-k\Pr[R_*\ge r+1]}\\
&=e^{(n\ln2-k)\Pr[R_*\ge r+1]}2^{-n\Pr[R_*\ge r+1]}\\
&\le e^{(n\ln2-k)^+}2^{-n\Pr[R_*\ge r+1]},
\end{align*}
where the first inequality uses \(1-u\le e^{-u}\).

The exponential factor in the last line is independent of \(r\) and is at least one. We can therefore substitute this bound into \eqref{eq:maximum-threshold-identity}, including the term \(2^{-n}\), and apply Lemma~\ref{lem:threshold-sum} to \(R_*\). This gives
\begin{equation}\label{eq:max-inhomogeneous}
\E2^{-M}
\le(8+4\sqrt2)e^{(n\ln2-k)^+}2^{-H(R_*)-\E R_*}.
\end{equation}

It remains to relate \(H(R_*)+\E R_*\) to the average entropy of the original distributions. Since \(R_*\) has the average of the laws of \(R_1,\ldots,R_k\), concavity of Shannon entropy gives
\[
H(R_*)\ge\frac1k\sum_{i=1}^kH(R_i).
\]
Also, \(\E R_*=k^{-1}\sum_i\E R_i\). Combining these relations with Lemma~\ref{lem:layer-entropy}, applied separately to each \(Q_i\), yields
\[
H(R_*)+\E R_*
\ge\frac1k\sum_{i=1}^k\bigl(H(R_i)+\E R_i\bigr)
\ge\frac1k\sum_{i=1}^kH(Q_i)-1
=\overline H-1.
\]
Substituting this inequality into \eqref{eq:max-inhomogeneous} gives
\[
\E2^{-M}
\le2(8+4\sqrt2)e^{(n\ln2-k)^+}2^{-\overline H}.
\]
Together with \eqref{eq:layer-average-bound} and \(C_0=2(8+4\sqrt2)\), this proves \eqref{eq:intro-flattening}.
\end{proof}

\subsection{Compound-Poisson consequence}

We next apply the same layer argument when the number of factors is Poisson distributed, as required by the prediction bound in Section~\ref{sec:prediction}. The main difference is that the Poisson generating function gives an exact expression for the probability that every selected scale is below a given threshold.

Recall from Section~\ref{sec:preliminaries} that \(P_\lambda\) denotes the distribution of the product of \(K_\lambda\sim\Pois(\lambda)\) independent random variables with common distribution \(Q\), where the count is independent of the factors.

\begin{corollary}[Compound-Poisson convolution bound]\label{cor:poisson-flattening}
Let \(Q\) be a distribution with finite support of size \(s\ge2\) on a discrete group, let \(n=\lfloor\log_2s\rfloor\), and let \(K_\lambda\sim\Pois(\lambda)\), where \(\lambda\ge0\). Then
\begin{equation}\label{eq:poisson-annealed}
\lVert P_\lambda\rVert_\infty
\le\E\lVert Q^{*K_\lambda}\rVert_\infty
\le C_0e^{(n\ln2-\lambda)^+}2^{-H(Q)}.
\end{equation}
\end{corollary}

\begin{proof}
For each factor, independently select a uniform layer of \(Q\) according to Lemma~\ref{lem:layer-decomposition}, and let \(R_i\) be its scale. These scales are independent and have a common distribution, which we denote by \(R\). Let \(M\) be the maximum of the first \(K_\lambda\) scales, with \(M=0\) when \(K_\lambda=0\).

Conditional on \(K_\lambda=k\), the product has distribution \(Q^{*k}\). For \(k\ge1\), \eqref{eq:conditional-layer-bound} and averaging over the selected layers give
\[
\lVert Q^{*k}\rVert_\infty
\le\E[2^{-M}\mid K_\lambda=k].
\]
The same inequality holds for \(k=0\), since \(Q^{*0}=\delta_e\) and \(M=0\). Averaging over \(K_\lambda\), and using the fact that \(P_\lambda\) is a mixture of the convolution powers, gives
\begin{equation}\label{eq:poisson-layer-average}
\lVert P_\lambda\rVert_\infty
\le\E\lVert Q^{*K_\lambda}\rVert_\infty
\le\E2^{-M}.
\end{equation}

We now bound \(\E2^{-M}\). For \(0\le r<n\), conditional on \(K_\lambda=k\), the event \(M\le r\) occurs precisely when all \(k\) selected scales are at most \(r\). Since the scales are independent of \(K_\lambda\), the Poisson generating function \eqref{eq:poisson-pgf} in Appendix~\ref{sec:standard-tools} gives
\[
\Pr[M\le r]
=\E\bigl[(1-\Pr[R\ge r+1])^{K_\lambda}\bigr]
=e^{-\lambda\Pr[R\ge r+1]}
\le e^{(n\ln2-\lambda)^+}2^{-n\Pr[R\ge r+1]}.
\]
Substituting this bound into the threshold identity \eqref{eq:maximum-threshold-identity} and applying Lemma~\ref{lem:threshold-sum} to \(R\) yields
\[
\E2^{-M}
\le(8+4\sqrt2)e^{(n\ln2-\lambda)^+}2^{-H(R)-\E R}.
\]

Finally, Lemma~\ref{lem:layer-entropy} gives \(H(R)+\E R\ge H(Q)-1\), and therefore
\[
\E2^{-M}
\le2(8+4\sqrt2)e^{(n\ln2-\lambda)^+}2^{-H(Q)}.
\]
Combining this with \eqref{eq:poisson-layer-average} and \(C_0=2(8+4\sqrt2)\) proves \eqref{eq:poisson-annealed}.
\end{proof}

In particular, choosing \(\lambda=\ln s\) ensures \(n\ln2\le\lambda\), so the exponential factor disappears and
\[
\lVert P_\lambda\rVert_\infty\le C_0 2^{-H(Q)}.
\]
This is the compound-Poisson bound used in the inner-product prediction argument.

\section{Inner-Product Prediction over Finite Fields}\label{sec:prediction}

We now use the compound-Poisson convolution bound to establish a prediction inequality for inner products over finite fields. Let \(q\) be a prime power, \(m\ge1\), and \(\mathcal V=\F_q^m\). Consider independent random column vectors \(\boldsymbol X,\boldsymbol Y\in\mathcal V\). A decoder observes \(\boldsymbol Y\), but not \(\boldsymbol X\), and must predict \(\boldsymbol X^{\mathsf T}\boldsymbol Y\). We will show that a small prediction error requires the two input distributions to have sufficiently large combined KL divergence from uniformity. This is the inequality needed when the inputs are drawn from the posterior distributions of a communication protocol.

For a fixed \(\boldsymbol y\in\mathcal V\), an optimal decoder chooses a most likely value of \(\boldsymbol X^{\mathsf T}\boldsymbol y\), succeeding with probability \(\max_{a\in\F_q}\Pr[\boldsymbol X^{\mathsf T}\boldsymbol y=a]\). Uniform guessing succeeds with probability \(1/q\). We normalize the advantage of optimal prediction over uniform guessing by
\begin{equation}\label{eq:modal-advantage}
\rho_{P_{\boldsymbol X}}(\boldsymbol y)
=\frac{q\max_{a\in\F_q}\Pr[\boldsymbol X^{\mathsf T}\boldsymbol y=a]-1}{q-1}.
\end{equation}
Thus, \(0\le\rho_{P_{\boldsymbol X}}(\boldsymbol y)\le1\). The advantage is zero when the inner product is uniformly distributed over \(\F_q\), and one when its value is determined by \(\boldsymbol y\). In particular, \(\rho_{P_{\boldsymbol X}}(\boldsymbol0)=1\), since the inner product with \(\boldsymbol0\) is always zero. Recall that \(L_q=\log_2q\) and \(c_0=\log_2(16+8\sqrt2)\).

\begin{theorem}[Optimal prediction bound]\label{thm:modal-prediction}
Let \(\boldsymbol X\) and \(\boldsymbol Y\) be independent random vectors on \(\mathcal V=\F_q^m\). Then
\begin{equation}\label{eq:modal-prediction}
D(P_{\boldsymbol X}\Vert U_{\mathcal V})+D(P_{\boldsymbol Y}\Vert U_{\mathcal V})
\ge mL_q\,\E\rho_{P_{\boldsymbol X}}(\boldsymbol Y)-c_0.
\end{equation}
\end{theorem}

We prove the theorem by applying the Gibbs variational formula to a score derived from an optimal prediction rule. The resulting bound involves an exponential moment, which we will control using the compound-Poisson convolution bound.

Let \(\phi:\mathcal V\to\F_q\) be an optimal prediction rule, choosing a most likely value of \(\boldsymbol X^{\mathsf T}\boldsymbol y\) for each \(\boldsymbol y\). For a fixed \(\boldsymbol x\in\mathcal V\), define
\begin{equation}\label{eq:prediction-score}
f_\phi(\boldsymbol x)
=1-\frac q{q-1}\Pr_{\boldsymbol Y}
[\boldsymbol x^{\mathsf T}\boldsymbol Y\ne\phi(\boldsymbol Y)].
\end{equation}
This score assigns weight \(1\) to a correct prediction and \(-1/(q-1)\) to an incorrect one, averaged over \(\boldsymbol Y\). By independence of \(\boldsymbol X\) and \(\boldsymbol Y\), averaging the score over \(\boldsymbol X\) gives
\begin{equation}\label{eq:prediction-score-average}
\E_{\boldsymbol X}f_\phi(\boldsymbol X)
=\frac{q\Pr[\boldsymbol X^{\mathsf T}\boldsymbol Y=\phi(\boldsymbol Y)]-1}{q-1}
=\E_{\boldsymbol Y}\rho_{P_{\boldsymbol X}}(\boldsymbol Y).
\end{equation}
The last equality uses the optimality of \(\phi\) at each value of \(\boldsymbol Y\).

The Gibbs variational formula~\cite[Lemma~4.10]{vanHandel2016} bounds the expectation of a function under \(P_{\boldsymbol X}\) by its divergence from a reference distribution and an exponential moment under that reference distribution. Applying it with reference distribution \(U_{\mathcal V}\) gives
\begin{equation}\label{eq:prediction-gibbs}
mL_q\,\E_{\boldsymbol X}f_\phi(\boldsymbol X)
\le D(P_{\boldsymbol X}\Vert U_{\mathcal V})
+\log_2\E_{\boldsymbol U\sim U_{\mathcal V}}2^{mL_q f_\phi(\boldsymbol U)}.
\end{equation}
By \eqref{eq:prediction-score-average}, the left-hand side is the prediction advantage appearing in Theorem~\ref{thm:modal-prediction}. Therefore, it remains to choose an optimal rule \(\phi\) for which
\begin{equation}\label{eq:modal-mgf}
\log_2\E_{\boldsymbol U\sim U_{\mathcal V}}2^{mL_q f_\phi(\boldsymbol U)}
\le D(P_{\boldsymbol Y}\Vert U_{\mathcal V})+c_0.
\end{equation}
The next lemma establishes precisely this bound.

\begin{lemma}[Exponential moment of an optimal prediction rule]\label{lem:prediction-exponential-moment}
Let \(\boldsymbol X\) and \(\boldsymbol Y\) be independent random vectors on \(\mathcal V\). There exists an optimal prediction rule \(\phi:\mathcal V\to\F_q\) satisfying \eqref{eq:modal-mgf} with the score \(f_\phi\) defined in \eqref{eq:prediction-score}.
\end{lemma}

\begin{proof}
We will encode \(\boldsymbol Y\) and its predicted inner product in a random vector \(\boldsymbol Z\), and use a Poisson sum of independent copies of \(\boldsymbol Z\) to bound the exponential moment. To apply the convolution bound, the encoding must have support size at most \(q^m\) and entropy at least \(H(\boldsymbol Y)\). We first construct an optimal prediction rule whose output scales with its input.

For each one-dimensional subspace of \(\mathcal V\), choose a nonzero representative \(\boldsymbol y_0\) and a most likely value \(a_{\boldsymbol y_0}\) of \(\boldsymbol X^{\mathsf T}\boldsymbol y_0\). Define \(\phi(\boldsymbol0)=0\) and
\[
\phi(t\boldsymbol y_0)=t a_{\boldsymbol y_0},
\qquad t\in\F_q^*.
\]
Since \(\boldsymbol X^{\mathsf T}(t\boldsymbol y_0)=t\boldsymbol X^{\mathsf T}\boldsymbol y_0\), multiplication by a nonzero scalar permutes the possible values of the inner product without changing their probabilities. Hence, \(\phi\) is optimal at every \(\boldsymbol y\) and satisfies
\[
\phi(t\boldsymbol y)=t\phi(\boldsymbol y)
\qquad\text{for all }t\in\F_q^*.
\]

Let \(T\) be uniform on \(\F_q^*\), independently of \(\boldsymbol Y\), and define
\[
\boldsymbol Z=T(\boldsymbol Y,-\phi(\boldsymbol Y))\in\F_q^{m+1}.
\]
Since \(\phi(T\boldsymbol Y)=T\phi(\boldsymbol Y)\), we have
\[
\boldsymbol Z=(T\boldsymbol Y,-\phi(T\boldsymbol Y)).
\]
Thus, \(\boldsymbol Z\) is determined by its first \(m\) coordinates, and its support has size at most \(q^m\). Moreover, for each fixed value of \(T\), the first \(m\) coordinates determine \(\boldsymbol Y\), since \(T\ne0\). Consequently,
\begin{equation}\label{eq:Z-entropy}
H(\boldsymbol Z)\ge H(\boldsymbol Z\mid T)=H(\boldsymbol Y).
\end{equation}
The random scaling also gives an important symmetry: for every \(c\in\F_q^*\), the random vectors \(c\boldsymbol Z\) and \(\boldsymbol Z\) have the same distribution, because \(cT\) is again uniform on \(\F_q^*\).

We next relate the distribution of \(\boldsymbol Z\) to the prediction score \(f_\phi\) using Fourier analysis over \(\F_q\). Fix a nontrivial additive character \(\psi\) of \(\F_q\), so that \(\psi(a+b)=\psi(a)\psi(b)\). For a random vector \(\boldsymbol W\) on \(\F_q^d\), define its Fourier transform by
\[
\widehat P_{\boldsymbol W}(\boldsymbol u)
=\E\psi(\boldsymbol u^{\mathsf T}\boldsymbol W),
\qquad \boldsymbol u\in\F_q^d.
\]
The character construction and Fourier inversion formula are given in Appendix~\ref{sec:finite-field-characters}.

Character orthogonality implies that, for any \(b\in\F_q\),
\[
\E_T\psi(Tb)=
\begin{cases}
1, & b=0,\\
-\dfrac1{q-1}, & b\ne0.
\end{cases}
\]
Indeed, when \(b\ne0\), the product \(Tb\) is uniform on \(\F_q^*\), and the sum of a nontrivial additive character over \(\F_q\) is zero. Applying this identity with \(b=\boldsymbol x^{\mathsf T}\boldsymbol Y-\phi(\boldsymbol Y)\), we obtain
\begin{equation}\label{eq:Zhat-modal}
\widehat P_{\boldsymbol Z}(\boldsymbol x,1)
=\E\psi\bigl(T(\boldsymbol x^{\mathsf T}\boldsymbol Y-\phi(\boldsymbol Y))\bigr)
=f_\phi(\boldsymbol x).
\end{equation}
Thus, the Fourier coefficient of \(\boldsymbol Z\) at \((\boldsymbol x,1)\) equals the prediction score for \(\boldsymbol x\).

We now use a Poisson number of copies to produce the exponential appearing in \eqref{eq:modal-mgf}. Let \(\boldsymbol Z_1,\boldsymbol Z_2,\ldots\) be independent copies of \(\boldsymbol Z\), and let \(K\sim\Pois(m\ln q)\) be independent of them. Define the sum
\[
\boldsymbol S=\sum_{i=1}^K\boldsymbol Z_i\in\F_q^{m+1},
\]
with the empty sum equal to zero. By independence and the Poisson generating function \eqref{eq:poisson-pgf} in Appendix~\ref{sec:standard-tools},
\[
\widehat P_{\boldsymbol S}(\boldsymbol x,1)
=\E_K\bigl[\widehat P_{\boldsymbol Z}(\boldsymbol x,1)^K\bigr]
=e^{m\ln q(f_\phi(\boldsymbol x)-1)}
=q^{-m}2^{mL_qf_\phi(\boldsymbol x)}.
\]
The choice of Poisson mean serves two purposes. It produces the exponential moment with the factor \(q^{-m}\), and, since \(|\supp P_{\boldsymbol Z}|\le q^m\), it is large enough to apply the compound-Poisson convolution bound.

It remains to express the average of these Fourier coefficients as a difference of two point probabilities of \(\boldsymbol S\). The scaling symmetry of \(\boldsymbol Z\) is inherited by \(\boldsymbol S\), since each summand has the same distribution after multiplication by any \(c\in\F_q^*\). In terms of Fourier transforms, this implies
\[
\widehat P_{\boldsymbol S}(\boldsymbol x,\tau)
=\widehat P_{\boldsymbol S}(\tau^{-1}\boldsymbol x,1)
\qquad\text{for every }\tau\in\F_q^*.
\]

Fix \(a\in\F_q^*\). Fourier inversion gives
\[
P_{\boldsymbol S}(\boldsymbol0,0)-P_{\boldsymbol S}(\boldsymbol0,a)
=q^{-(m+1)}
\sum_{\tau\in\F_q^*}
(1-\psi(-\tau a))
\sum_{\boldsymbol x\in\mathcal V}
\widehat P_{\boldsymbol S}(\boldsymbol x,\tau).
\]
The terms with \(\tau=0\) cancel in the difference. For every nonzero \(\tau\), the scaling identity above shows that the inner sum equals \(\sum_{\boldsymbol x\in\mathcal V}\widehat P_{\boldsymbol S}(\boldsymbol x,1)\). Character orthogonality also gives
\[
\sum_{\tau\in\F_q^*}(1-\psi(-\tau a))=q.
\]
Combining these identities and substituting the expression for \(\widehat P_{\boldsymbol S}(\boldsymbol x,1)\), we obtain
\begin{equation}\label{eq:prediction-exponential-moment}
P_{\boldsymbol S}(\boldsymbol0,0)-P_{\boldsymbol S}(\boldsymbol0,a)
=q^{-m}\E_{\boldsymbol U\sim U_{\mathcal V}}2^{mL_q f_\phi(\boldsymbol U)}
\le\|P_{\boldsymbol S}\|_\infty.
\end{equation}
The last inequality follows because \(P_{\boldsymbol S}(\boldsymbol0,a)\ge0\). Appendix~\ref{sec:prediction-fourier-details} records the Fourier calculation.

We can now apply the convolution bound. If the support of \(\boldsymbol Z\) contains at least two points, the Poisson mean \(m\ln q\) is at least \(\ln|\supp P_{\boldsymbol Z}|\), so Corollary~\ref{cor:poisson-flattening} gives
\[
\|P_{\boldsymbol S}\|_\infty
\le C_0 2^{-H(\boldsymbol Z)}
\le C_0 2^{-H(\boldsymbol Y)},
\]
where the second inequality uses \eqref{eq:Z-entropy}. If the support of \(\boldsymbol Z\) is a singleton, then \eqref{eq:Z-entropy} implies \(H(\boldsymbol Y)=0\), and the same bound follows from \(\|P_{\boldsymbol S}\|_\infty\le1\).

Combining this bound with \eqref{eq:prediction-exponential-moment}, multiplying by \(q^m\), and taking base-two logarithms yields
\[
\log_2\E_{\boldsymbol U\sim U_{\mathcal V}}2^{mL_q f_\phi(\boldsymbol U)}
\le mL_q-H(\boldsymbol Y)+c_0
=D(P_{\boldsymbol Y}\Vert U_{\mathcal V})+c_0.
\]
This is precisely \eqref{eq:modal-mgf}.
\end{proof}

\begin{proof}[Proof of Theorem~\ref{thm:modal-prediction}]
Choose the optimal prediction rule supplied by Lemma~\ref{lem:prediction-exponential-moment}. Substituting \eqref{eq:modal-mgf} into the Gibbs bound \eqref{eq:prediction-gibbs}, and then using \eqref{eq:prediction-score-average}, gives \eqref{eq:modal-prediction}.
\end{proof}

Theorem~\ref{thm:modal-prediction} concerns optimal prediction. For information complexity, however, we need a bound for the decoder induced by a communication protocol, which need not be optimal. Since no decoder can have a higher success probability than an optimal one, the theorem immediately gives the following inequality.

\begin{corollary}[Prediction bound for any decoder]\label{cor:decoder-aware}
Let \(\boldsymbol X\) and \(\boldsymbol Y\) be independent random vectors on \(\mathcal V=\F_q^m\), and let \(z:\mathcal V\to\F_q\). If
\[
\eta=\Pr[z(\boldsymbol Y)\ne\boldsymbol X^{\mathsf T}\boldsymbol Y],
\]
then
\begin{equation}\label{eq:decoder-aware}
D(P_{\boldsymbol X}\Vert U_{\mathcal V})+D(P_{\boldsymbol Y}\Vert U_{\mathcal V})
\ge mL_q\left(1-\frac q{q-1}\eta\right)-c_0.
\end{equation}
\end{corollary}

\begin{proof}
For each \(\boldsymbol y\), the probability of correctly predicting \(\boldsymbol X^{\mathsf T}\boldsymbol y\) using \(z(\boldsymbol y)\) is at most the success probability of an optimal prediction. Averaging over \(\boldsymbol Y\), the optimal success probability is therefore at least \(1-\eta\). By \eqref{eq:modal-advantage},
\[
\E_{\boldsymbol Y}\rho_{P_{\boldsymbol X}}(\boldsymbol Y)
\ge\frac{q(1-\eta)-1}{q-1}
=1-\frac q{q-1}\eta.
\]
Substituting this inequality into Theorem~\ref{thm:modal-prediction} proves \eqref{eq:decoder-aware}.
\end{proof}

\subsection{Exact prediction at zero error}\label{sec:exact-prediction-details}

When the prediction error is zero, we can obtain a stronger bound without the additive constant \(c_0\). In this case, every input pair in the supports of the two distributions must produce the same inner product as the decoder's output for the corresponding \(\boldsymbol Y\). This imposes an orthogonality constraint on the supports, which gives the entropy bound directly.

\begin{lemma}[Exact inner-product prediction]\label{lem:exact-inner-product-prediction}
Let \(\boldsymbol X\) and \(\boldsymbol Y\) be independent random vectors on \(\mathcal V=\F_q^m\), and suppose \(z:\mathcal V\to\F_q\) satisfies
\[
\Pr[z(\boldsymbol Y)=\boldsymbol X^{\mathsf T}\boldsymbol Y]=1.
\]
Then
\begin{equation}\label{eq:exact-inner-product-prediction}
D(P_{\boldsymbol X}\Vert U_{\mathcal V})
+D(P_{\boldsymbol Y}\Vert U_{\mathcal V})\ge mL_q.
\end{equation}
\end{lemma}

\begin{proof}
Choose any \(\boldsymbol x_0\in\supp P_{\boldsymbol X}\), and let
\[
\mathcal W=\operatorname{span}\{\boldsymbol x-\boldsymbol x_0:\boldsymbol x\in\supp P_{\boldsymbol X}\}.
\]
By construction, \(\supp P_{\boldsymbol X}\subseteq\boldsymbol x_0+\mathcal W\).

Since \(\boldsymbol X\) and \(\boldsymbol Y\) are independent, every pair \((\boldsymbol x,\boldsymbol y)\in\supp P_{\boldsymbol X}\times\supp P_{\boldsymbol Y}\) has positive probability. The zero-error assumption therefore implies
\[
\boldsymbol x^{\mathsf T}\boldsymbol y=z(\boldsymbol y)
=\boldsymbol x_0^{\mathsf T}\boldsymbol y
\]
for every such pair. Hence, \((\boldsymbol x-\boldsymbol x_0)^{\mathsf T}\boldsymbol y=0\) for every \(\boldsymbol x\in\supp P_{\boldsymbol X}\) and \(\boldsymbol y\in\supp P_{\boldsymbol Y}\). By the definition of \(\mathcal W\), this gives \(\supp P_{\boldsymbol Y}\subseteq\mathcal W^\perp\).

Writing \(r=\dim(\mathcal W)\), the supports of \(\boldsymbol X\) and \(\boldsymbol Y\) are therefore contained in sets of sizes \(q^r\) and \(q^{m-r}\), respectively. Consequently,
\[
H(\boldsymbol X)\le rL_q,
\qquad
H(\boldsymbol Y)\le(m-r)L_q.
\]
Their sum is at most \(mL_q\). Since
\[
D(P_{\boldsymbol X}\Vert U_{\mathcal V})
+D(P_{\boldsymbol Y}\Vert U_{\mathcal V})
=2mL_q-H(\boldsymbol X)-H(\boldsymbol Y),
\]
equation~\eqref{eq:exact-inner-product-prediction} follows.
\end{proof}

\section{Information Complexity}\label{sec:bilinear-ic}

We now prove Theorem~\ref{thm:matrix-multiplication}, which determines the information complexity of computing \(\boldsymbol X\boldsymbol Y^{\mathsf T}\) to within a universal additive constant under independent uniform inputs. The lower bound combines scalar prediction with an entropy inequality for random linear projections. After proving the matrix theorem, we derive the rank-dependent bounds for scalar bilinear forms and the corresponding transcript-determined upper bound.

\subsection{Transcript posteriors and protocol cost}

To prove lower bounds on information complexity, we first bound the information cost of an arbitrary protocol before taking the infimum over protocols. Recall that \(\IC_\mu(\Pi)\) denotes the information cost of a specified protocol, defined in \eqref{eq:internal-information}, whereas \(\IC_\mu^{\mathrm B,\mathrm{av}}(f,\varepsilon)\) and \(\IC_\mu^{\mathrm B,\mathrm{pw}}(f,\varepsilon)\) are infima over Bob-only protocols satisfying the corresponding error constraints. The superscript \(\mathrm{tr}\) denotes the transcript-determined output convention.

Let \(X\in\mathcal X\) and \(Y\in\mathcal Y\) be independent uniform inputs to a protocol with transcript \(\Pi\). Conditioning on a transcript generally changes the distributions of the inputs, but it preserves their independence. Moreover, the information cost of the protocol equals the average divergence of these conditional distributions, called the posteriors, from the original uniform distributions. We state these properties in the following lemma.

\begin{lemma}[Posterior product structure]\label{lem:posterior-product}
Suppose the input prior is \(U_{\mathcal X\times\mathcal Y}\). For each transcript value \(\pi\) with \(\Pr[\Pi=\pi]>0\),
\[
P_{X,Y\mid\Pi=\pi}
=P_{X\mid\pi}\times P_{Y\mid\pi},
\]
where \(P_{X\mid\pi}=P_{X\mid\Pi=\pi}\) and \(P_{Y\mid\pi}=P_{Y\mid\Pi=\pi}\). Moreover,
\begin{equation}\label{eq:posterior-information}
\IC_{U_{\mathcal X\times\mathcal Y}}(\Pi)
=\E_\Pi\bigl[D(P_{X\mid\Pi}\Vert U_{\mathcal X})
+D(P_{Y\mid\Pi}\Vert U_{\mathcal Y})\bigr].
\end{equation}
\end{lemma}

The independence assertion follows from the factorization of protocol likelihoods: for each fixed transcript \(\pi\), its probability conditional on the inputs can be written as \(a_\pi(x)b_\pi(y)\), where \(a_\pi\) depends only on Alice's input and \(b_\pi\) only on Bob's input. Multiplying this likelihood by the product prior and normalizing therefore gives a product posterior. The information-cost identity then follows by expressing the two conditional mutual informations in \eqref{eq:internal-information} as expected posterior divergences. The full proof is given in Appendix~\ref{sec:local-decoding-posteriors}.

For Bob-only protocols, Lemma~\ref{lem:deterministic-decoding} also allows us to replace Bob's final output by a deterministic decoder \(z(Y,\Pi)\), without increasing average error or changing the information cost. Once a transcript \(\pi\) is fixed, this becomes a predictor \(z(Y,\pi)\) based only on Bob's input, while the posterior inputs remain independent. We can therefore apply the prediction bounds from Section~\ref{sec:prediction} at each transcript and then average the resulting divergence bounds using \eqref{eq:posterior-information}.

\subsection{Entropy under random linear projections}

In the matrix multiplication problem, scalar tests of the output lead to inner products between linear projections of the two inputs. To apply the prediction bounds to these projected inputs, we need to relate their divergences from uniformity to those of the original inputs. The following theorem shows that, when a linear projection reduces a matrix from \(t\) rows to \(d\) rows, its expected divergence from uniformity is at most \(d/t\) times the original divergence. Here the projection acts by left multiplication, applying the same linear map to each column of the input matrix. The proof uses Shearer's inequality, which relates the entropy of a collection of random variables to the entropies of its subsets, together with a random invertible change of basis~\cite{ChungGrahamFranklShearer1986,MadimanTetali2010}.

\begin{theorem}[Rank-normalized entropy projection]\label{thm:rank-normalized-projection}
Let \(\boldsymbol Z\) be an arbitrary random matrix on \(\F_q^{t\times m}\), and independently let \(\boldsymbol T\) be uniform over the full-row-rank matrices in \(\F_q^{d\times t}\), where \(1\le d\le t\). Then
\begin{equation}\label{eq:rank-projection-entropy}
H(\boldsymbol T\boldsymbol Z\mid\boldsymbol T)
\ge \frac dt H(\boldsymbol Z)
\end{equation}
and
\begin{equation}\label{eq:rank-projection-divergence}
\E_{\boldsymbol T}D\!\left(
P_{\boldsymbol T\boldsymbol Z\mid\boldsymbol T}
\middle\Vert U_{\F_q^{d\times m}}\right)
\le \frac dt D\!\left(P_{\boldsymbol Z}\middle\Vert U_{\F_q^{t\times m}}\right).
\end{equation}
\end{theorem}

\begin{proof}
We represent a uniform full-row-rank projection by first applying a random invertible change of basis and then selecting \(d\) rows. Let \(\boldsymbol G\) be uniform on \(\operatorname{GL}_t(q)\), the set of invertible \(t\times t\) matrices over \(\F_q\), and independent of \(\boldsymbol Z\). Independently choose a uniform \(d\)-element subset \(S\subseteq\{1,\ldots,t\}\), and let \(\boldsymbol E_S\) select the rows indexed by \(S\) in increasing order.

For each fixed \(S\), every ordered collection of \(d\) linearly independent row vectors can be completed to an invertible \(t\times t\) matrix in the same number of ways. Since \(\boldsymbol G\) is uniform over all invertible matrices, \(\boldsymbol E_S\boldsymbol G\) is uniform over the full-row-rank matrices in \(\F_q^{d\times t}\). It therefore has the same distribution as \(\boldsymbol T\).

Now fix an invertible matrix \(\boldsymbol g\). The \(t\) rows of \(\boldsymbol g\boldsymbol Z\) form \(t\) jointly distributed random vectors in \(\F_q^m\). Shearer's inequality states that the average entropy of a uniformly chosen subset of \(d\) such vectors is at least \(d/t\) times their joint entropy. Applying it to these row vectors gives
\[
\E_S H\bigl((\boldsymbol g\boldsymbol Z)_S\bigr)
\ge\frac dt H(\boldsymbol g\boldsymbol Z)
=\frac dt H(\boldsymbol Z),
\]
where the equality follows because left multiplication by \(\boldsymbol g\) is invertible and hence preserves entropy. Averaging over \(\boldsymbol G\), and using the fact that \(\boldsymbol E_S\boldsymbol G\) has the same distribution as \(\boldsymbol T\), proves \eqref{eq:rank-projection-entropy}.

To obtain the divergence bound, recall that divergence from the uniform distribution equals the difference between the maximum possible entropy and the actual entropy. Since the projected matrix has \(d\times m\) entries over \(\F_q\),
\begin{align*}
\E_{\boldsymbol T}D\!\left(
P_{\boldsymbol T\boldsymbol Z\mid\boldsymbol T}
\middle\Vert U_{\F_q^{d\times m}}\right)
&=dmL_q-H(\boldsymbol T\boldsymbol Z\mid\boldsymbol T)\\
&\le dmL_q-\frac dt H(\boldsymbol Z)\\
&=\frac dt D\!\left(
P_{\boldsymbol Z}\middle\Vert U_{\F_q^{t\times m}}\right).
\end{align*}
This proves \eqref{eq:rank-projection-divergence}.
\end{proof}

\subsection{Proof of the matrix information bound}\label{sec:matrix-ic}

We now prove Theorem~\ref{thm:matrix-multiplication}. The main idea is to test the matrix output using a random linear functional, thereby reducing matrix multiplication to an inner-product prediction problem. The rank of the test determines the dimension of the projected inputs. We use Theorem~\ref{thm:rank-normalized-projection} to relate their divergences to those of the original inputs. The following bound on the reciprocal rank ensures that the additive loss remains independent of the matrix dimensions.

\begin{lemma}[Reciprocal-rank bound]\label{lem:reciprocal-rank}
For every prime power \(q\) and \(1\le t\le s\),
\begin{equation}\label{eq:kappa-upper}
\kappa_{q,t,s}\le2.
\end{equation}
\end{lemma}

The proof is given in Appendix~\ref{sec:matrix-details}.

\begin{proof}[Proof of Theorem~\ref{thm:matrix-multiplication}]

Fix a Bob-only protocol with transcript \(\Pi\) and average error at most \(\varepsilon\)  under \(U_{\F_q^{t\times m}\times\F_q^{s\times m}}\). By Lemma~\ref{lem:deterministic-decoding}, we can replace Bob's output by a deterministic decoder \(\widehat{\boldsymbol M}(\boldsymbol Y,\Pi)\) without increasing the average error or changing the information cost. For each positive-probability transcript \(\pi\), the posterior inputs \(\boldsymbol X\) and \(\boldsymbol Y\) remain independent by Lemma~\ref{lem:posterior-product}. Write
\[
\Delta_{\boldsymbol X}(\pi)=D(P_{\boldsymbol X\mid\pi}\Vert U_{\F_q^{t\times m}}),
\qquad
\Delta_{\boldsymbol Y}(\pi)=D(P_{\boldsymbol Y\mid\pi}\Vert U_{\F_q^{s\times m}}),
\]
and let \(\eta_\pi\) be the conditional probability that \(\widehat{\boldsymbol M}(\boldsymbol Y,\pi)\ne\boldsymbol X\boldsymbol Y^{\mathsf T}\). We first derive a lower bound on the sum of these two divergences in terms of \(\eta_\pi\).

Choose a uniform nonzero matrix \(\boldsymbol\Lambda\in\F_q^{t\times s}\), independently of the inputs and transcript. We test Bob's output by comparing the scalar \(\langle\boldsymbol\Lambda,\widehat{\boldsymbol M}(\boldsymbol Y,\pi)\rangle_{\mathrm F}\) with \(\langle\boldsymbol\Lambda,\boldsymbol X\boldsymbol Y^{\mathsf T}\rangle_{\mathrm F}\), where \(\langle\cdot,\cdot\rangle_{\mathrm F}\) denotes the entrywise inner product. Let \(\eta_{\pi,\boldsymbol\Lambda}\) be the conditional probability that these two scalars differ.

The error of this scalar test has a simple relation to the matrix error. For any fixed nonzero error matrix \(\boldsymbol E\), the map
\[
\boldsymbol\Lambda\longmapsto
\langle\boldsymbol\Lambda,\boldsymbol E\rangle_{\mathrm F}
\]
is a nonzero linear functional on the \(ts\)-dimensional vector space \(\F_q^{t\times s}\). Its kernel has \(q^{ts-1}\) elements, so exactly \(q^{ts-1}(q-1)\) of the \(q^{ts}-1\) nonzero test matrices detect the error. If the matrix output is correct, every scalar test is also correct. Averaging over the test matrices therefore gives
\begin{equation}\label{eq:average-matrix-character-error}
\E_{\boldsymbol\Lambda}\eta_{\pi,\boldsymbol\Lambda}
=\eta_\pi\frac{q^{ts-1}(q-1)}{q^{ts}-1}.
\end{equation}
Thus, the average scalar error determines the matrix error up to an explicit factor.

To apply the inner-product prediction bound, we express the tested scalar as an inner product of projected inputs. Condition on \(\rank\boldsymbol\Lambda=d\), where \(1\le d\le t\), and choose uniformly a factorization
\begin{equation}\label{eq:uniform-rank-factorization}
\boldsymbol\Lambda=\boldsymbol L^{\mathsf T}\boldsymbol R,
\qquad
\boldsymbol L\in\F_q^{d\times t},
\quad
\boldsymbol R\in\F_q^{d\times s},
\end{equation}
with both factors of full row rank. Every rank-\(d\) matrix has exactly \(|\operatorname{GL}_d(q)|\) such factorizations, since a choice of basis for its column space determines \(\boldsymbol L^{\mathsf T}\) and then uniquely determines \(\boldsymbol R\). Consequently, conditional on rank \(d\), the pair \((\boldsymbol L,\boldsymbol R)\) is uniform over all pairs of full-row-rank matrices of the indicated sizes. In particular, \(\boldsymbol L\) and \(\boldsymbol R\) are independent and uniformly distributed over their respective sets of full-row-rank matrices.

By the factorization,
\begin{equation}\label{eq:matrix-character-factorization}
\langle\boldsymbol\Lambda,\boldsymbol X\boldsymbol Y^{\mathsf T}\rangle_{\mathrm F}
=\langle\boldsymbol L\boldsymbol X,\boldsymbol R\boldsymbol Y\rangle_{\mathrm F}.
\end{equation}
The matrices \(\boldsymbol L\boldsymbol X\) and \(\boldsymbol R\boldsymbol Y\), viewed as vectors in \(\F_q^{dm}\), remain independent under the posterior distributions. There is, however, one point to address before applying Corollary~\ref{cor:decoder-aware}: Bob's tested output may depend on his entire input \(\boldsymbol Y\), rather than only on \(\boldsymbol R\boldsymbol Y\).

For fixed \(\pi,\boldsymbol L,\boldsymbol R\), the target \(\langle\boldsymbol L\boldsymbol X,\boldsymbol R\boldsymbol Y\rangle_{\mathrm F}\) depends on Bob's input only through \(\boldsymbol R\boldsymbol Y\). Conditional on this projection, the remaining randomness in \(\boldsymbol Y\) is independent of \(\boldsymbol L\boldsymbol X\), because the posterior inputs are independent. Thus, Bob's tested output can be regarded as a randomized predictor based on \(\boldsymbol R\boldsymbol Y\). Replacing it, for each projected input, by a most likely value of the target cannot increase the prediction error. This reduction is formalized in Lemma~\ref{lem:bilinear-prediction} of Appendix~\ref{sec:bilinear-prediction-details}.

We can therefore apply Corollary~\ref{cor:decoder-aware} to the projected inputs in dimension \(dm\). For each fixed factorization, their divergences satisfy
\[
\begin{aligned}
&D(P_{\boldsymbol L\boldsymbol X\mid\pi}\Vert U_{\F_q^{d\times m}})
+D(P_{\boldsymbol R\boldsymbol Y\mid\pi}\Vert U_{\F_q^{d\times m}})\\
&\qquad\ge dmL_q\left(1-\frac q{q-1}\eta_{\pi,\boldsymbol\Lambda}\right)-c_0.
\end{aligned}
\]
Now average over the factorizations conditional on rank \(d\). Since \(\boldsymbol L\) and \(\boldsymbol R\) are independent uniform full-row-rank matrices, Theorem~\ref{thm:rank-normalized-projection} bounds the expected divergences on the left by \(d\Delta_{\boldsymbol X}(\pi)/t\) and \(d\Delta_{\boldsymbol Y}(\pi)/s\), respectively. Dividing the resulting inequality by \(d\), we obtain
\begin{equation}\label{eq:projected-matrix-prediction}
\frac{\Delta_{\boldsymbol X}(\pi)}t+\frac{\Delta_{\boldsymbol Y}(\pi)}s
\ge mL_q\left(1-\frac q{q-1}\E[\eta_{\pi,\boldsymbol\Lambda}\mid\rank\boldsymbol\Lambda=d]\right)-\frac{c_0}d.
\end{equation}

We next average over the rank of \(\boldsymbol\Lambda\). By \eqref{eq:average-matrix-character-error},
\[
\frac q{q-1}\E_{\boldsymbol\Lambda}\eta_{\pi,\boldsymbol\Lambda}
=\frac{q^{ts}}{q^{ts}-1}\eta_\pi.
\]
Moreover, the definition of \(\kappa_{q,t,s}\) in \eqref{eq:kappa-matrix-multiplication} gives
\[
\E_{\boldsymbol\Lambda}\frac1{\rank\boldsymbol\Lambda}
=\frac{\kappa_{q,t,s}}t.
\]
Substituting these identities into the average of \eqref{eq:projected-matrix-prediction} yields
\begin{equation}\label{eq:weighted-matrix-posterior}
\frac{\Delta_{\boldsymbol X}(\pi)}t+\frac{\Delta_{\boldsymbol Y}(\pi)}s
\ge mL_q\left(1-\frac{q^{ts}}{q^{ts}-1}\eta_\pi\right)-\frac{c_0}t\kappa_{q,t,s}.
\end{equation}

Finally, since \(t\le s\) and both divergences are nonnegative,
\[
\Delta_{\boldsymbol X}(\pi)+\Delta_{\boldsymbol Y}(\pi)
\ge\Delta_{\boldsymbol X}(\pi)+\frac ts\Delta_{\boldsymbol Y}(\pi).
\]
Multiplying \eqref{eq:weighted-matrix-posterior} by \(t\) therefore gives a lower bound on the sum of the original posterior divergences. Averaging over transcripts, using \(\E_\Pi\eta_\Pi\le\varepsilon\), and applying \eqref{eq:posterior-information}, we obtain
\begin{equation}\label{eq:matrix-posterior-lower}
\IC_{U_{\F_q^{t\times m}\times\F_q^{s\times m}}}(\Pi)\ge mtL_q\left(1-\frac{q^{ts}}{q^{ts}-1}\varepsilon\right)-c_0\kappa_{q,t,s}.
\end{equation}
Together with nonnegativity of information cost, this proves the average-error lower bound in Theorem~\ref{thm:matrix-multiplication}. It also applies to protocols with pointwise error at most \(\varepsilon\), and to transcript-determined output, since these protocols are subject to stronger constraints.

For the upper bound, use public randomness to choose between exact computation and uniform guessing. With probability \(\alpha_{q,t,s}(\varepsilon)\), Alice sends \(\boldsymbol X\), allowing Bob to compute the matrix product exactly. Otherwise, Bob outputs an independent uniform matrix in \(\F_q^{t\times s}\). On every fixed input pair, the latter branch has error probability \(1-q^{-ts}\), so the overall pointwise error is
\[
(1-\alpha_{q,t,s}(\varepsilon))(1-q^{-ts})\le\varepsilon.
\]
Only the exact branch reveals information about the inputs. Since \(\boldsymbol X\) is uniform on \(\F_q^{t\times m}\), the protocol's information cost is \(\alpha_{q,t,s}(\varepsilon)mtL_q\), as required.

At zero error, we can remove the additive loss in the lower bound. Every scalar test then has zero prediction error, so Lemma~\ref{lem:exact-inner-product-prediction} replaces Corollary~\ref{cor:decoder-aware} in the argument above. The resulting version of \eqref{eq:projected-matrix-prediction} has no term \(c_0/d\). Averaging over test ranks and transcripts, and multiplying by \(t\), gives \(\IC_{U_{\F_q^{t\times m}\times\F_q^{s\times m}}}(\Pi)\ge mtL_q\). Sending \(\boldsymbol X\) achieves this cost, proving the exact zero-error statement.
\end{proof}

When \(t=s=1\), every nonzero test matrix has rank one, and hence \(\kappa_{q,1,1}=1\). Theorem~\ref{thm:matrix-multiplication} therefore gives the inner-product bound with additive loss \(c_0\), including Corollary~\ref{cor:intro-binary}.

\subsection{Scalar bilinear forms}

We next extend the information-complexity bounds to scalar bilinear forms. Let \(\boldsymbol A\in\F_q^{n\times m}\) have rank \(r\ge1\), and define
\[
B_{\boldsymbol A}(\boldsymbol x,\boldsymbol y)
=\boldsymbol x^{\mathsf T}\boldsymbol A\boldsymbol y\in\F_q.
\]
Choose a rank factorization \(\boldsymbol A=\boldsymbol A_{\mathrm L}^{\mathsf T}\boldsymbol A_{\mathrm R}\), where \(\boldsymbol A_{\mathrm L}\in\F_q^{r\times n}\) and \(\boldsymbol A_{\mathrm R}\in\F_q^{r\times m}\) both have full row rank. Then
\[
B_{\boldsymbol A}(\boldsymbol x,\boldsymbol y)
=(\boldsymbol A_{\mathrm L}\boldsymbol x)^{\mathsf T}
(\boldsymbol A_{\mathrm R}\boldsymbol y).
\]
Thus, the bilinear form is an inner product of two \(r\)-dimensional projections. We use this representation to transfer the inner-product lower bound to \(B_{\boldsymbol A}\). For the upper bound, Alice needs to communicate only her projection. We also consider the additional information cost when the output must be determined by the transcript.

For \(0\le\varepsilon\le1\), write
\begin{equation}\label{eq:alpha-q}
\alpha_q(\varepsilon)=\left(1-\frac q{q-1}\varepsilon\right)^+.
\end{equation}

\begin{corollary}[Information complexity of a scalar bilinear form]\label{thm:scalar-bilinear-ic}
For \(0\le\varepsilon\le1\),
\begin{equation}\label{eq:scalar-bob-bounds}
\begin{aligned}
\bigl[rL_q\alpha_q(\varepsilon)-c_0\bigr]^+
&\le\IC_{U_{\F_q^n\times\F_q^m}}^{\mathrm B,\mathrm{av}}(B_{\boldsymbol A},\varepsilon)\\
&\le\IC_{U_{\F_q^n\times\F_q^m}}^{\mathrm B,\mathrm{pw}}(B_{\boldsymbol A},\varepsilon)
\le rL_q\alpha_q(\varepsilon).
\end{aligned}
\end{equation}
For transcript-determined output, the same average-error lower bound holds. A pointwise-error upper bound is
\begin{equation}\label{eq:scalar-transcript-upper}
\IC_{U_{\F_q^n\times\F_q^m}}^{\mathrm{tr},\mathrm{pw}}(B_{\boldsymbol A},\varepsilon)
\le\alpha_q(\varepsilon)\bigl(r+1-q^{-r}\bigr)L_q.
\end{equation}
For Bob-only output, the zero-error information complexity is
\begin{equation}\label{eq:scalar-zero-error}
\IC_{U_{\F_q^n\times\F_q^m}}^{\mathrm B}(B_{\boldsymbol A},0)=rL_q.
\end{equation}
\end{corollary}

\begin{proof}

We first prove the lower bound by reducing the \(r\)-dimensional inner-product problem to \(B_{\boldsymbol A}\). Consider an arbitrary protocol for \(B_{\boldsymbol A}\) under the product-uniform input distribution, and let \(\Pi\) denote its transcript.

Suppose Alice and Bob instead receive independent uniform vectors \(\boldsymbol u,\boldsymbol v\in\F_q^r\), and wish to compute \(\boldsymbol u^{\mathsf T}\boldsymbol v\). Alice privately samples \(\boldsymbol X\) uniformly from the fiber
\[
\{\boldsymbol x\in\F_q^n:\boldsymbol A_{\mathrm L}\boldsymbol x=\boldsymbol u\},
\]
while Bob independently samples \(\boldsymbol Y\) uniformly from
\[
\{\boldsymbol y\in\F_q^m:\boldsymbol A_{\mathrm R}\boldsymbol y=\boldsymbol v\}.
\]
They then run the original protocol on \(\boldsymbol X,\boldsymbol Y\). Because both projection maps have full row rank, they are surjective and all their fibers have equal size. Consequently, the lifted inputs \(\boldsymbol X\) and \(\boldsymbol Y\) are independent and uniform on \(\F_q^n\) and \(\F_q^m\), respectively. Moreover,
\[
B_{\boldsymbol A}(\boldsymbol X,\boldsymbol Y)
=(\boldsymbol A_{\mathrm L}\boldsymbol X)^{\mathsf T}
(\boldsymbol A_{\mathrm R}\boldsymbol Y)
=\boldsymbol u^{\mathsf T}\boldsymbol v.
\]
Thus, the simulated protocol computes the \(r\)-dimensional inner product with the same average error as the original protocol.

We next compare their information costs. Under the simulation, \(\boldsymbol u=\boldsymbol A_{\mathrm L}\boldsymbol X\) and \(\boldsymbol v=\boldsymbol A_{\mathrm R}\boldsymbol Y\). Since the input distributions are products, Lemma~\ref{lem:posterior-product} expresses the information cost of each protocol as the sum of the two unconditional mutual informations with its transcript. By data processing,
\[
\begin{aligned}
&I(\boldsymbol A_{\mathrm L}\boldsymbol X;\Pi)
+I(\boldsymbol A_{\mathrm R}\boldsymbol Y;\Pi)\\
&\qquad\le I(\boldsymbol X;\Pi)+I(\boldsymbol Y;\Pi).
\end{aligned}
\]
Therefore, the simulated inner-product protocol has no larger information cost than the original protocol for \(B_{\boldsymbol A}\). Applying Theorem~\ref{thm:matrix-multiplication} with \(t=s=1\) and inner-product dimension \(r\), for which \(\kappa_{q,1,1}=1\) and \(\alpha_{q,1,1}(\varepsilon)=\alpha_q(\varepsilon)\), proves the lower bound in \eqref{eq:scalar-bob-bounds}. The same reduction and the exact zero-error bound in that theorem give \(\IC_{U_{\F_q^n\times\F_q^m}}^{\mathrm B}(B_{\boldsymbol A},0)\ge rL_q\). The average-error lower bound also applies to transcript-determined protocols, since they are a special case of Bob-only protocols.

For the Bob-only upper bound, use public randomness to choose between exact computation and uniform guessing. With probability \(\alpha_q(\varepsilon)\), Alice sends \(\boldsymbol A_{\mathrm L}\boldsymbol X\), allowing Bob to compute
\[
(\boldsymbol A_{\mathrm L}\boldsymbol X)^{\mathsf T}
(\boldsymbol A_{\mathrm R}\boldsymbol Y)
\]
exactly. Otherwise, Bob outputs an independent uniform element of \(\F_q\). The pointwise error is
\[
(1-\alpha_q(\varepsilon))(1-q^{-1})\le\varepsilon.
\]
Since \(\boldsymbol A_{\mathrm L}\boldsymbol X\) is uniform on \(\F_q^r\), the information cost of the exact branch is \(rL_q\), while the guessing branch reveals no information about the inputs. The overall information cost is therefore \(\alpha_q(\varepsilon)rL_q\), proving the upper bound in \eqref{eq:scalar-bob-bounds}. At zero error, this protocol attains \(rL_q\), establishing \eqref{eq:scalar-zero-error}.

For transcript-determined output, we modify the exact branch so that Bob also sends the computed scalar to Alice. The guessing branch uses a public uniform field element as the output, so the output is determined by the transcript in both branches. It remains to calculate the information cost of Bob's additional message in the exact branch.

Conditional on Alice's projection \(\boldsymbol A_{\mathrm L}\boldsymbol X\), the scalar output is zero if that projection is zero. Otherwise, it is uniform on \(\F_q\), because \(\boldsymbol A_{\mathrm R}\boldsymbol Y\) is independent and uniform on \(\F_q^r\), and its inner product with any fixed nonzero vector is uniform on \(\F_q\). Since \(\boldsymbol A_{\mathrm L}\boldsymbol X\) is uniform, it is zero with probability \(q^{-r}\). Hence,
\[
H\bigl(B_{\boldsymbol A}(\boldsymbol X,\boldsymbol Y)
\mid\boldsymbol A_{\mathrm L}\boldsymbol X\bigr)
=(1-q^{-r})L_q.
\]
This is precisely the additional information about Bob's input revealed by his message, after Alice's projection is known. The exact branch therefore has information cost
\[
rL_q+(1-q^{-r})L_q
=\bigl(r+1-q^{-r}\bigr)L_q.
\]
Multiplying by the probability \(\alpha_q(\varepsilon)\) of taking this branch proves \eqref{eq:scalar-transcript-upper}.
\end{proof}

\section{Conclusion}\label{sec:conclusion}

We have determined the information complexity of binary inner product under independent uniform inputs to within a universal additive constant throughout the error range \(0\le\varepsilon\le1/2\). For protocols in which Bob produces the output, the information complexity is \((1-2\varepsilon)m\) up to this additive loss. The upper bound is achieved by mixing exact computation with uniform guessing, while the lower bound applies even when error is averaged over the inputs. The same approach gives bounds for matrix multiplication over finite fields and scalar bilinear forms, with an additive gap independent of the matrix dimensions and field size. At zero error, we obtain the exact information complexity.

The main ingredient is a convolution theorem that bounds the min-entropy of a product of independent random variables in terms of their average Shannon entropy, with only a universal constant loss. Its compound-Poisson form controls the exponential moment needed for the prediction inequality, which in turn yields information lower bounds by averaging over protocol transcripts. For matrix multiplication, random linear tests and entropy bounds for projections allow us to apply this prediction inequality without accumulating losses across matrix entries. The convolution theorem applies to arbitrary discrete groups, and the leading coefficient of its logarithmic threshold is optimal. Open questions include determining the optimal additive loss in the convolution bound, the exact information complexity at positive error, and whether a similar entropy conversion holds in the presence of side information.

\clearpage
\appendix

\section{Standard Poisson and Finite-Field Fourier Identities}\label{sec:standard-tools}

\subsection{Poisson counts}

Let \(K\sim\Pois(\lambda)\). Its probability generating function is
\begin{equation}\label{eq:poisson-pgf}
\E t^K=e^{\lambda(t-1)}.
\end{equation}
This identity holds for every \(t\in\mathbb C\), with \(t^0=1\). In particular, suppose each of the \(K\) points is independently marked with probability \(p\). Conditional on \(K\), the probability that no point is marked is \((1-p)^K\). Averaging over \(K\) and applying \eqref{eq:poisson-pgf} gives
\[
\Pr[\text{no point is marked}]=e^{-\lambda p}.
\]

\subsection{Finite-field characters}\label{sec:finite-field-characters}

Write \(q=p^\ell\), where \(p\) is prime and \(\ell\ge1\). An additive character of \(\F_q\) is a function \(\psi:\F_q\to\mathbb C\) satisfying \(|\psi(a)|=1\) and \(\psi(a+b)=\psi(a)\psi(b)\). It is nontrivial if it is not identically one. We use the character
\[
\psi(a)=e^{2\pi i\operatorname{Tr}_{\F_q/\F_p}(a)/p},
\]
where
\[
\operatorname{Tr}_{\F_q/\F_p}(a)=\sum_{j=0}^{\ell-1}a^{p^j}\in\F_p
\]
is the field trace, and its value in the exponent is represented by an integer in \(\{0,\ldots,p-1\}\). When \(\ell=1\), the trace is the identity and \(\psi(a)=e^{2\pi i a/q}\). The field trace is additive and surjective onto \(\F_p\), so \(\psi\) is a nontrivial additive character with \(|\psi(a)|=1\).

For \(\mathcal V=\F_q^m\), character orthogonality gives
\begin{equation}\label{eq:character-orthogonality}
\frac1{|\mathcal V|}\sum_{\boldsymbol x\in\mathcal V}
\psi(\boldsymbol x^{\mathsf T}\boldsymbol y)
=\1_{\{\boldsymbol y=\boldsymbol0\}}.
\end{equation}
Indeed, the sum equals one after normalization when \(\boldsymbol y=\boldsymbol0\). Otherwise, the map \(\boldsymbol x\mapsto\psi(\boldsymbol x^{\mathsf T}\boldsymbol y)\) is a nontrivial additive character on \(\mathcal V\), whose sum is zero. The identities below hold for any nontrivial additive character.

For a random vector \(\boldsymbol W\) on \(\F_q^d\), define the Fourier transform of its distribution by
\[
\widehat P_{\boldsymbol W}(\boldsymbol u)
=\E\psi(\boldsymbol u^{\mathsf T}\boldsymbol W),
\qquad \boldsymbol u\in\F_q^d.
\]
By character orthogonality, the distribution can be recovered from its Fourier transform:
\begin{equation}\label{eq:finite-field-fourier-inversion}
P_{\boldsymbol W}(\boldsymbol w)
=q^{-d}\sum_{\boldsymbol u\in\F_q^d}
\widehat P_{\boldsymbol W}(\boldsymbol u)
\psi(-\boldsymbol u^{\mathsf T}\boldsymbol w).
\end{equation}

Let \(\boldsymbol W_1,\boldsymbol W_2,\ldots\) be independent copies of \(\boldsymbol W\). For every integer \(k\ge1\), independence and the character identity give
\[
\widehat P_{\boldsymbol W_1+\cdots+\boldsymbol W_k}(\boldsymbol u)
=\widehat P_{\boldsymbol W}(\boldsymbol u)^k.
\]
Now let \(K\sim\Pois(\lambda)\) be independent of these copies, and let \(P_\lambda\) denote the distribution of their random sum \(\sum_{i=1}^K\boldsymbol W_i\), with the empty sum equal to zero. Conditioning on \(K\) and applying \eqref{eq:poisson-pgf} yields
\begin{equation}\label{eq:finite-field-poisson-fourier}
\widehat P_\lambda(\boldsymbol u)
=e^{\lambda(\widehat P_{\boldsymbol W}(\boldsymbol u)-1)}.
\end{equation}

\section{The Directed-Cycle and Threshold Estimates}\label{sec:threshold-details}

We prove the threshold-sum bound used in Section~\ref{sec:flattening}. The main ingredient is an entropy inequality for expected distances on a directed cycle. We first establish this inequality and then apply it to the distribution of the layer scale.

For \(n\ge1\), identify \(\Z_n\) with \(\{0,\ldots,n-1\}\) and define the clockwise distance from \(s\) to \(r\) by
\[
c(s,r)=(r-s)\bmod n\in\{0,\ldots,n-1\}.
\]
When \(n=1\), the only distance is zero.

\begin{lemma}[Directed-cycle entropy bound]\label{lem:cycle}
Let \(P\) be a distribution on \(\Z_n\), where \(n\ge1\), and let \(S\sim P\). If \(d_r=\E c(S,r)\), then
\begin{equation}\label{eq:cycle-bound}
2^{H(P)}\sum_{r\in\Z_n}2^{-d_r}\le6+4\sqrt2.
\end{equation}
\end{lemma}

\begin{proof}

We bound the sum \(z=\sum_r2^{-d_r}\) by interpreting its terms as probabilities. Define a distribution \(\nu\) on \(\Z_n\) by
\[
\nu(r)=\frac{2^{-d_r}}z.
\]
Let \(T\sim\nu\) be independent of \(S\), and let \(C=c(S,T)=(T-S)\bmod n\) be the clockwise distance from \(S\) to \(T\). By the definition of \(d_r\),
\[
\E[C\mid T=r]=d_r.
\]
Moreover, the choice of \(\nu\) gives
\[
H(T)=\sum_r\nu(r)(d_r+\log_2z)
=\E C+\log_2z.
\]
Thus,
\[
\log_2z=H(T)-\E C.
\]

We next relate the entropies of \(S\) and \(T\) to that of their distance \(C\). For each fixed \(T=r\), the map \(S\mapsto(r-S)\bmod n\) is a bijection on \(\Z_n\). Since \(S\) and \(T\) are independent, this gives \(H(C\mid T)=H(S)\). Similarly, fixing \(S\) gives \(H(C\mid S)=H(T)\). Conditioning cannot increase entropy, so
\[
H(C)\ge H(C\mid T)=H(S),
\qquad
H(C)\ge H(C\mid S)=H(T).
\]
Consequently,
\[
H(P)+\log_2z
=H(S)+H(T)-\E C
\le2H(C)-\E C.
\]
It remains to bound the last expression by a universal constant.

Since \(C\) takes nonnegative integer values, we compare its distribution with the geometric distribution
\[
\gamma(j)=(1-2^{-1/2})2^{-j/2},
\qquad j=0,1,\ldots.
\]
The nonnegativity of KL divergence gives
\[
D(P_C\Vert\gamma)
=-H(C)-\log_2(1-2^{-1/2})+\frac12\E C
\ge0.
\]
Rearranging,
\[
H(C)-\frac12\E C
\le-\log_2(1-2^{-1/2}).
\]
Therefore,
\[
2H(C)-\E C
\le-2\log_2(1-2^{-1/2})
=\log_2(6+4\sqrt2).
\]
Combining this with the preceding bound on \(H(P)+\log_2z\) yields
\[
z\le(6+4\sqrt2)2^{-H(P)},
\]
which proves \eqref{eq:cycle-bound}.
\end{proof}

\begin{proof}[Proof of Lemma~\ref{lem:threshold-sum}]

Let \(R\) be supported on \(\{0,\ldots,n\}\), as in the statement of Lemma~\ref{lem:threshold-sum}. We first express the exponents in the threshold sum as expected clockwise distances, so that Lemma~\ref{lem:cycle} can be applied.

Let \(\overline R=R\bmod n\), identifying the two endpoints \(0\) and \(n\) on \(\Z_n\). For each \(0\le r<n\), the clockwise distance from \(\overline R\) to \(r\) equals \(r-R\) when \(R\le r\), and \(r-R+n\) when \(R\ge r+1\). Taking expectations gives
\begin{equation}\label{eq:cycle-identity}
\E[(r-\overline R)\bmod n]
=r-\E R+n\Pr[R\ge r+1].
\end{equation}
Write \(d_r\) for this expected distance. Applying Lemma~\ref{lem:cycle} to \(\overline R\), we obtain
\[
\sum_{r=0}^{n-1}2^{-d_r}
\le(6+4\sqrt2)2^{-H(\overline R)}.
\]

To relate this bound to the entropy of \(R\), observe that \(\overline R\) determines \(R\) except when \(\overline R=0\), in which case \(R\) can be either \(0\) or \(n\). Thus, \(H(R\mid\overline R)\le1\). Since \(\overline R\) is a function of \(R\), the entropy chain rule gives
\[
H(\overline R)=H(R)-H(R\mid\overline R)\ge H(R)-1.
\]
Consequently,
\begin{equation}\label{eq:threshold-cycle-part}
\sum_{r=0}^{n-1}2^{-d_r}
\le2(6+4\sqrt2)2^{-H(R)}.
\end{equation}
By \eqref{eq:cycle-identity}, we have
\[
r+n\Pr[R\ge r+1]=\E R+d_r.
\]
Hence, \eqref{eq:threshold-cycle-part} bounds the sum in Lemma~\ref{lem:threshold-sum}:
\[
\begin{aligned}
\frac12\sum_{r=0}^{n-1}2^{-r-n\Pr[R\ge r+1]}
&=\frac12\cdot2^{-\E R}\sum_{r=0}^{n-1}2^{-d_r}\\
&\le(6+4\sqrt2)2^{-H(R)-\E R}.
\end{aligned}
\]

It remains to control the endpoint term \(2^{-n}\) by the same entropy expression. Let \(W=n-R\), and compare its distribution with the geometric distribution
\[
\gamma_0(w)=2^{-w-1},
\qquad w=0,1,\ldots.
\]
By nonnegativity of KL divergence,
\[
D(P_W\Vert\gamma_0)
=-H(W)+\E W+1\ge0.
\]
Since \(W=n-R\) is a bijective function of \(R\), we have \(H(W)=H(R)\) and \(\E W=n-\E R\). Therefore,
\[
H(R)\le n-\E R+1,
\]
which implies
\begin{equation}\label{eq:threshold-endpoint}
2^{-n}\le2\cdot2^{-H(R)-\E R}.
\end{equation}

Combining the bounds for the sum and the endpoint term gives
\[
2^{-n}+\frac12\sum_{r=0}^{n-1}2^{-r-n\Pr[R\ge r+1]}
\le(8+4\sqrt2)2^{-H(R)-\E R}.
\]
This proves Lemma~\ref{lem:threshold-sum}.

\end{proof}

\section{Sharpness}\label{sec:sharpness}

We use a mixture of a point mass and a uniform distribution to establish three limitations of the convolution bounds: the leading coefficient of the logarithmic threshold cannot be reduced, the coefficient of Shannon entropy cannot exceed one, and the additive loss cannot vanish uniformly at the Poisson threshold.

Let \(G_s=\Z_s\) for \(s\ge2\), and let \(U_{G_s}\) be the uniform distribution on \(G_s\). For \(0<p<1\), define
\[
Q_{p,s}=(1-p)\delta_0+pU_{G_s}.
\]
To sample from \(Q_{p,s}\), first choose the point-mass component with probability \(1-p\) or the uniform component with probability \(p\), then draw a value from the chosen component. Let \(P_{p,s,\lambda}\) denote the compound-Poisson distribution with rate \(\lambda\), formed from independent variables with common distribution \(Q_{p,s}\). We also write
\[
h_2(p)=-p\log_2p-(1-p)\log_2(1-p)
\]
for the binary entropy function.

\begin{proposition}[Lazy-uniform obstruction]\label{prop:lazy-uniform}
For every integer \(j\ge0\) and every \(\lambda\ge0\),
\begin{align}
Q_{p,s}^{*j}
&=(1-p)^j\delta_0+\bigl(1-(1-p)^j\bigr)U_{G_s},\label{eq:lazy-deterministic}\\
P_{p,s,\lambda}
&=e^{-p\lambda}\delta_0+\bigl(1-e^{-p\lambda}\bigr)U_{G_s}.\label{eq:lazy-poisson}
\end{align}

Moreover, for fixed \(p\in(0,1)\),
\begin{equation}\label{eq:lazy-entropy}
H(Q_{p,s})=p\log_2s+h_2(p)+o(1)
\end{equation}
as \(s\to\infty\).
\end{proposition}

\begin{proof}

Sample each factor by independently choosing the point-mass component with probability \(1-p\) or the uniform component with probability \(p\). The convolution of \(U_{G_s}\) with any distribution on \(G_s\) is again \(U_{G_s}\). Thus, if at least one factor is drawn from the uniform component, the sum of all factors is uniform on \(G_s\). Otherwise, every factor equals zero, and so does their sum.

For \(j\) factors, the probability that no uniform component is selected is \((1-p)^j\). This gives \eqref{eq:lazy-deterministic}. When the number of factors is \(K_\lambda\sim\Pois(\lambda)\), the corresponding probability is
\[
\E(1-p)^{K_\lambda}=e^{-p\lambda}
\]
by \eqref{eq:poisson-pgf}, proving \eqref{eq:lazy-poisson}.

Finally, \(Q_{p,s}\) assigns probability \(1-p+p/s\) to zero and \(p/s\) to each nonzero element. Therefore,
\[
\begin{aligned}
H(Q_{p,s})
={}&-\left(1-p+\frac ps\right)\log_2\left(1-p+\frac ps\right)\\
&-(s-1)\frac ps\log_2\frac ps.
\end{aligned}
\]
For fixed \(p\), this equals \(p\log_2s+h_2(p)+o(1)\) as \(s\to\infty\), proving \eqref{eq:lazy-entropy}.
\end{proof}

We first consider the number of factors required for a constant entropy loss. Fix \(0<c<1\) and \(0<p<1\), and take the Poisson rate \(\lambda=c\ln s\). By \eqref{eq:lazy-poisson}, the largest point probability is attained at zero and equals
\[
\lVert P_{p,s,c\ln s}\rVert_\infty
=s^{-cp}+\frac{1-s^{-cp}}s
=s^{-cp}(1+o(1)).
\]
Here the last equality uses \(cp<1\). Consequently,
\[
H_\infty(P_{p,s,c\ln s})
=cp\log_2s+o(1).
\]
Together with \eqref{eq:lazy-entropy}, this gives
\[
H(Q_{p,s})-H_\infty(P_{p,s,c\ln s})
=(1-c)p\log_2s+h_2(p)+o(1).
\]
The difference diverges as \(s\to\infty\). Thus, a Poisson rate \(c\ln s\) with fixed \(c<1\) cannot guarantee a universal constant loss relative to the input Shannon entropy.

A similar obstruction holds when the number of factors is fixed. Let \(j=\lfloor c\ln s\rfloor\). By \eqref{eq:lazy-deterministic},
\[
\lVert Q_{p,s}^{*j}\rVert_\infty\ge(1-p)^j.
\]
Since \(j\le c\ln s\) and \(0<1-p<1\), we have
\[
(1-p)^j\ge s^{c\ln(1-p)}.
\]
Combining this with \eqref{eq:lazy-entropy} yields
\[
\lVert Q_{p,s}^{*j}\rVert_\infty\,2^{H(Q_{p,s})}
\ge 2^{h_2(p)+o(1)}s^{p+c\ln(1-p)}.
\]
For sufficiently small fixed \(p>0\), the exponent is positive, since
\[
p+c\ln(1-p)=(1-c)p+O(p^2)
\]
as \(p\to0\). The right-hand side therefore diverges as \(s\to\infty\). Hence, a fixed number of factors with leading coefficient \(c<1\) in the logarithmic threshold cannot guarantee a uniform constant loss in min-entropy.

Finally, consider the Poisson rate \(\lambda=\ln s\), at which the convolution bound guarantees a constant entropy loss. Equation~\eqref{eq:lazy-poisson} gives
\[
\lVert P_{p,s,\ln s}\rVert_\infty
=s^{-p}+\frac{1-s^{-p}}s
=s^{-p}(1+o(1)),
\]
and hence
\[
H_\infty(P_{p,s,\ln s})=p\log_2s+o(1).
\]
For any fixed \(\delta>0\), we therefore have
\[
(1+\delta)H(Q_{p,s})-H_\infty(P_{p,s,\ln s})
=\delta p\log_2s+(1+\delta)h_2(p)+o(1),
\]
which diverges as \(s\to\infty\). Thus, the coefficient of \(H(Q_{p,s})\) in the min-entropy lower bound cannot be increased above one. Equivalently, there is no universal bound
\[
\lVert P_{p,s,\ln s}\rVert_\infty
\le C2^{-(1+\delta)H(Q_{p,s})}
\]
with \(C\) independent of \(s\).

The same calculation also gives
\[
H(Q_{p,s})-H_\infty(P_{p,s,\ln s})
=h_2(p)+o(1).
\]
Taking \(p=1/2\), for which \(h_2(p)=1\), shows that a universal additive loss at this Poisson rate must be at least one bit. Equivalently, the multiplicative constant in the corresponding largest-point-probability bound must be at least \(2\).

\section{Prediction Identities and Reduction to Rank}\label{sec:prediction-details}

We provide the Fourier calculation used in the inner-product prediction bound and prove the reduction from scalar bilinear prediction to inner-product prediction.

\subsection{Fourier details of the prediction moment}\label{sec:prediction-fourier-details}

We verify the identity \eqref{eq:prediction-exponential-moment} used in Lemma~\ref{lem:prediction-exponential-moment}. Recall the optimal prediction rule \(\phi\), the random vector \(\boldsymbol Z=T(\boldsymbol Y,-\phi(\boldsymbol Y))\), and the Poisson sum \(\boldsymbol S\) from the proof of that lemma.

We first relate the Fourier transform of \(\boldsymbol Z\) to the prediction score. Since \(\psi\) is a nontrivial additive character, character orthogonality gives
\[
\sum_{t\in\F_q}\psi(tb)=q\,\1_{\{b=0\}}
\]
for every \(b\in\F_q\). Removing the term \(t=0\) and recalling that \(T\) is uniform on \(\F_q^*\), we obtain
\[
\E_T\psi(Tb)=\frac{q\,\1_{\{b=0\}}-1}{q-1}.
\]
Applying this identity with \(b=\boldsymbol x^{\mathsf T}\boldsymbol Y-\phi(\boldsymbol Y)\), and then averaging over \(\boldsymbol Y\), gives
\[
\begin{aligned}
\widehat P_{\boldsymbol Z}(\boldsymbol x,1)
&=\E\psi\bigl(T(\boldsymbol x^{\mathsf T}\boldsymbol Y-\phi(\boldsymbol Y))\bigr)\\
&=1-\frac q{q-1}\Pr[\boldsymbol x^{\mathsf T}\boldsymbol Y\ne\phi(\boldsymbol Y)]\\
&=f_\phi(\boldsymbol x),
\end{aligned}
\]
as stated in \eqref{eq:Zhat-modal}.

Next, let \(K\sim\Pois(m\ln q)\) be independent of the copies of \(\boldsymbol Z\) defining \(\boldsymbol S\). By the compound-Poisson Fourier identity \eqref{eq:finite-field-poisson-fourier},
\[
\widehat P_{\boldsymbol S}(\boldsymbol x,\tau)
=e^{m\ln q(\widehat P_{\boldsymbol Z}(\boldsymbol x,\tau)-1)}.
\]
In particular,
\[
\widehat P_{\boldsymbol S}(\boldsymbol x,1)
=e^{m\ln q(f_\phi(\boldsymbol x)-1)}
=q^{-m}2^{mL_q f_\phi(\boldsymbol x)}.
\]

We now relate the average of these Fourier coefficients to point probabilities of \(\boldsymbol S\). Since \(T\) is uniform on \(\F_q^*\), the distributions of \(c\boldsymbol Z\) and \(\boldsymbol Z\) agree for every \(c\in\F_q^*\). This property also holds for \(\boldsymbol S\), because multiplying each summand by \(c\) does not change their joint distribution. Consequently, for every \(\tau\in\F_q^*\),
\[
\widehat P_{\boldsymbol S}(\boldsymbol x,\tau)
=\widehat P_{\boldsymbol S}(\tau^{-1}\boldsymbol x,1).
\]
Since multiplication by \(\tau^{-1}\) permutes \(\mathcal V\), summing over \(\boldsymbol x\) gives
\[
\sum_{\boldsymbol x\in\mathcal V}
\widehat P_{\boldsymbol S}(\boldsymbol x,\tau)
=
\sum_{\boldsymbol x\in\mathcal V}
\widehat P_{\boldsymbol S}(\boldsymbol x,1).
\]

Fix \(a\in\F_q^*\). Applying Fourier inversion \eqref{eq:finite-field-fourier-inversion} to the points \((\boldsymbol0,0)\) and \((\boldsymbol0,a)\), we obtain
\begin{align*}
&P_{\boldsymbol S}(\boldsymbol0,0)-P_{\boldsymbol S}(\boldsymbol0,a)\\
&\quad=q^{-(m+1)}
\sum_{\tau\in\F_q}\sum_{\boldsymbol x\in\mathcal V}
\widehat P_{\boldsymbol S}(\boldsymbol x,\tau)
(1-\psi(-\tau a))\\
&\quad=q^{-(m+1)}
\left(\sum_{\tau\ne0}(1-\psi(-\tau a))\right)
\sum_{\boldsymbol x\in\mathcal V}
\widehat P_{\boldsymbol S}(\boldsymbol x,1).
\end{align*}
The terms with \(\tau=0\) vanish because \(1-\psi(0)=0\). For every nonzero \(\tau\), the inner sum is the same by the scaling identity above. Finally, since \(a\ne0\), character orthogonality gives
\[
\sum_{\tau\ne0}\psi(-\tau a)=-1,
\]
so the coefficient in parentheses equals \(q\). Therefore,
\begin{equation}\label{eq:finite-field-slice}
P_{\boldsymbol S}(\boldsymbol0,0)-P_{\boldsymbol S}(\boldsymbol0,a)
=q^{-m}\sum_{\boldsymbol x\in\mathcal V}
\widehat P_{\boldsymbol S}(\boldsymbol x,1).
\end{equation}
Substituting the expression for \(\widehat P_{\boldsymbol S}(\boldsymbol x,1)\) and using the fact that \(\boldsymbol U\) is uniform on \(\mathcal V\), we obtain
\[
\begin{aligned}
P_{\boldsymbol S}(\boldsymbol0,0)-P_{\boldsymbol S}(\boldsymbol0,a)
&=q^{-2m}\sum_{\boldsymbol x\in\mathcal V}
2^{mL_q f_\phi(\boldsymbol x)}\\
&=q^{-m}\E_{\boldsymbol U}2^{mL_q f_\phi(\boldsymbol U)}.
\end{aligned}
\]
This is \eqref{eq:prediction-exponential-moment}.

\subsection{Bilinear prediction}\label{sec:bilinear-prediction-details}

Let \(\boldsymbol A\in\F_q^{n\times m}\) have rank \(r\ge1\), and write \(B_{\boldsymbol A}(\boldsymbol x,\boldsymbol y)=\boldsymbol x^{\mathsf T}\boldsymbol A\boldsymbol y\). We reduce prediction of this bilinear form to prediction of an inner product in \(\F_q^r\).

\begin{lemma}[Bilinear prediction]\label{lem:bilinear-prediction}
Let \(\boldsymbol X\) and \(\boldsymbol Y\) be independent random vectors on \(\F_q^n\) and \(\F_q^m\), with distributions \(P_{\boldsymbol X}\) and \(P_{\boldsymbol Y}\), respectively. If a decoder \(z(\boldsymbol Y)\) satisfies \(\Pr[z(\boldsymbol Y)\ne B_{\boldsymbol A}(\boldsymbol X,\boldsymbol Y)]=\eta\), then
\begin{equation}\label{eq:bilinear-prediction}
D(P_{\boldsymbol X}\Vert U_{\F_q^n})
+D(P_{\boldsymbol Y}\Vert U_{\F_q^m})
\ge rL_q\left(1-\frac q{q-1}\eta\right)-c_0.
\end{equation}
If \(\eta=0\), then the stronger bound
\begin{equation}\label{eq:exact-bilinear-prediction}
D(P_{\boldsymbol X}\Vert U_{\F_q^n})
+D(P_{\boldsymbol Y}\Vert U_{\F_q^m})\ge rL_q
\end{equation}
holds.
\end{lemma}

\begin{proof}

Choose a rank factorization
\[
\boldsymbol A=\boldsymbol A_{\mathrm L}^{\mathsf T}\boldsymbol A_{\mathrm R},
\qquad
\boldsymbol A_{\mathrm L}\in\F_q^{r\times n},
\quad
\boldsymbol A_{\mathrm R}\in\F_q^{r\times m},
\]
where both factors have full row rank. Then
\[
B_{\boldsymbol A}(\boldsymbol x,\boldsymbol y)
=(\boldsymbol A_{\mathrm L}\boldsymbol x)^{\mathsf T}
(\boldsymbol A_{\mathrm R}\boldsymbol y).
\]
Set \(\boldsymbol X'=\boldsymbol A_{\mathrm L}\boldsymbol X\) and \(\boldsymbol Y'=\boldsymbol A_{\mathrm R}\boldsymbol Y\). These projected inputs are independent, and the target becomes \(\boldsymbol X'^{\mathsf T}\boldsymbol Y'\).

The original decoder, however, may use the entire input \(\boldsymbol Y\), rather than only \(\boldsymbol Y'\). We show that this additional information cannot improve prediction of the projected inner product. Fix \(\boldsymbol v\in\supp P_{\boldsymbol Y'}\). Since \(\boldsymbol X'\) is independent of \(\boldsymbol Y\), it is also independent of the decoder output \(z(\boldsymbol Y)\). Consequently,
\[
\begin{aligned}
&\Pr[z(\boldsymbol Y)=\boldsymbol X'^{\mathsf T}\boldsymbol v
\mid\boldsymbol Y'=\boldsymbol v]\\
&\quad=\sum_{a\in\F_q}
\Pr[z(\boldsymbol Y)=a\mid\boldsymbol Y'=\boldsymbol v]
\Pr[\boldsymbol X'^{\mathsf T}\boldsymbol v=a]\\
&\quad\le\max_{a\in\F_q}
\Pr[\boldsymbol X'^{\mathsf T}\boldsymbol v=a].
\end{aligned}
\]
Thus, choosing a most likely value of \(\boldsymbol X'^{\mathsf T}\boldsymbol v\) gives a deterministic decoder based only on \(\boldsymbol v\), with no larger conditional error. Applying this choice for every \(\boldsymbol v\) in the support of \(\boldsymbol Y'\) gives a decoder of \(\boldsymbol X'^{\mathsf T}\boldsymbol Y'\) with error at most \(\eta\).

Corollary~\ref{cor:decoder-aware} therefore gives
\[
D(P_{\boldsymbol X'}\Vert U_{\F_q^r})
+D(P_{\boldsymbol Y'}\Vert U_{\F_q^r})
\ge rL_q\left(1-\frac q{q-1}\eta\right)-c_0.
\]
Since both projection maps have full row rank, they map the respective uniform input distributions to \(U_{\F_q^r}\). The data-processing inequality for KL divergence gives
\[
D(P_{\boldsymbol X'}\Vert U_{\F_q^r})
\le D(P_{\boldsymbol X}\Vert U_{\F_q^n}),
\qquad
D(P_{\boldsymbol Y'}\Vert U_{\F_q^r})
\le D(P_{\boldsymbol Y}\Vert U_{\F_q^m}).
\]
Combining these inequalities proves \eqref{eq:bilinear-prediction}.

If \(\eta=0\), the decoder based on \(\boldsymbol Y'\) also has zero error. Applying Lemma~\ref{lem:exact-inner-product-prediction} to the projected inputs and using the same data-processing inequalities proves \eqref{eq:exact-bilinear-prediction}.
\end{proof}

\section{Protocol Posteriors and Further Information Bounds}\label{sec:protocol-details}

We establish the protocol properties used to convert prediction bounds into information-complexity lower bounds. We then derive prior-free information-complexity bounds from the binary inner-product result.

\subsection{Local decoding and posterior factorization}\label{sec:local-decoding-posteriors}

We first show that, when only average error matters, Bob's private randomness is unnecessary for his final output once his input and the transcript are fixed.

\begin{lemma}[Deterministic local decoding]\label{lem:deterministic-decoding}
For every Bob-only protocol with a finite output alphabet and a fixed input distribution, there is a deterministic decoder \(z(Y,\Pi)\) with no larger average error and the same transcript distribution and information cost.
\end{lemma}

\begin{proof}

Let \(R_B\) denote Bob's private randomness, including any randomness used for his final output. Fix \(y,\pi\) with \(\Pr[Y=y,\Pi=\pi]>0\). Since the transcript records the public randomness and all exchanged messages, the protocol-tree likelihood factors according to Alice's input and Bob's input and private randomness. More precisely, there are nonnegative functions \(a_\pi(x)\) and \(b_\pi(y,b)\), with the distribution of Bob's private randomness absorbed into the latter, such that
\[
\Pr[X=x,R_B=b\mid Y=y,\Pi=\pi]
\propto
\Pr[X=x\mid Y=y]a_\pi(x)b_\pi(y,b).
\]
The factors involving \(x\) and \(b\) are separate. Hence, \(X\) and \(R_B\) are independent conditional on \((Y,\Pi)=(y,\pi)\).

The target value is determined by the input pair \((X,Y)\). Conditional on \((Y,\Pi)=(y,\pi)\), its distribution therefore does not depend on Bob's remaining private randomness. Choosing a most likely target value as \(z(y,\pi)\) minimizes the conditional prediction error. Applying this choice to every pair \((y,\pi)\) of positive probability and averaging over \((Y,\Pi)\) cannot increase the overall error. Since only the final decoder changes, the transcript distribution and information cost remain unchanged.
\end{proof}

We next prove that, under independent inputs, conditioning on a transcript preserves their independence. This property allows us to apply the prediction inequalities to the conditional input distributions.

\begin{proof}[Proof of Lemma~\ref{lem:posterior-product}]

For each full transcript \(\pi\), multiplying the message probabilities along the corresponding path in the protocol tree gives nonnegative functions \(a_\pi\) and \(b_\pi\) such that
\[
\Pr[\Pi=\pi\mid X=x,Y=y]=a_\pi(x)b_\pi(y).
\]
Under the product prior \(U_{\mathcal X}\times U_{\mathcal Y}\), Bayes' rule therefore gives
\[
P_{X,Y\mid\Pi=\pi}(x,y)
\propto
U_{\mathcal X}(x)a_\pi(x)\,
U_{\mathcal Y}(y)b_\pi(y).
\]
For every transcript of positive probability, the two factors can be normalized separately. Thus, the posterior distribution is a product, proving the first assertion of Lemma~\ref{lem:posterior-product}.

We now derive the information-cost identity. Since \(X\) and \(Y\) are independent under the prior,
\[
H(X\mid Y)=H(X),
\qquad
H(Y\mid X)=H(Y).
\]
Their conditional independence given \(\Pi\) similarly gives
\[
H(X\mid Y,\Pi)=H(X\mid\Pi),
\qquad
H(Y\mid X,\Pi)=H(Y\mid\Pi).
\]
Consequently,
\[
I(X;\Pi\mid Y)=I(X;\Pi),
\qquad
I(Y;\Pi\mid X)=I(Y;\Pi).
\]

Finally, mutual information can be expressed as the expected divergence of the posterior from the prior:
\[
I(X;\Pi)=\E_\Pi D(P_{X\mid\Pi}\Vert U_{\mathcal X}),
\qquad
I(Y;\Pi)=\E_\Pi D(P_{Y\mid\Pi}\Vert U_{\mathcal Y}).
\]
Substituting these identities into the definition of information cost \eqref{eq:internal-information} proves \eqref{eq:posterior-information}.
\end{proof}

\subsection{Prior-free information complexity}\label{sec:prior-free-details}

We now derive bounds that do not fix the input distribution in advance. Following \cite{DFHL2018}, let \(\IC(f,\varepsilon)\) be the maximum, over input distributions \(\mu\), of the infimum of the \(\mu\)-information cost over protocols with pointwise error at most \(\varepsilon\). Similarly, let \(\IC^D(f,\varepsilon)\) be the maximum over \(\mu\) of the infimum of the \(\mu\)-information cost over protocols with average error at most \(\varepsilon\) under \(\mu\). We use Bob-only output throughout this subsection.

\begin{corollary}[Prior-free inner product]\label{cor:prior-free}
For \(0\le\varepsilon\le1/2\),
\begin{equation}\label{eq:prior-free-bounds}
\bigl[(1-2\varepsilon)m-c_0\bigr]^+
\le\IC^D(\IP_m,\varepsilon)
\le\IC(\IP_m,\varepsilon)
\le(1-2\varepsilon)m.
\end{equation}
Consequently,
\[
0\le\IC(\IP_m,\varepsilon)-\IC^D(\IP_m,\varepsilon)\le c_0.
\]
\end{corollary}

\begin{proof}

For the lower bound, consider the product-uniform distribution \(U_{\F_2^m\times\F_2^m}\). Since \(\IC^D(\IP_m,\varepsilon)\) maximizes over all input distributions, it is at least the information complexity under \(U_{\F_2^m\times\F_2^m}\) with average error at most \(\varepsilon\). Corollary~\ref{cor:intro-binary} therefore gives
\[
\IC^D(\IP_m,\varepsilon)
\ge\bigl[(1-2\varepsilon)m-c_0\bigr]^+.
\]

For the upper bound, consider an arbitrary input distribution \(\mu\). With public probability \(1-2\varepsilon\), Alice sends \(\boldsymbol X\) and Bob computes the inner product exactly. Otherwise, Bob outputs a uniform random bit without communication. For every fixed input pair, the error probability is \(\varepsilon\), so the protocol satisfies the pointwise error constraint. Its information cost under \(\mu\) is
\[
(1-2\varepsilon)H(\boldsymbol X\mid\boldsymbol Y)
\le(1-2\varepsilon)m.
\]
This upper bound holds for every \(\mu\), and hence also after maximizing over input distributions.

Finally, for every fixed \(\mu\), a protocol satisfying the pointwise error constraint also satisfies the average-error constraint under \(\mu\). Thus, the infimum of the information cost under the average-error constraint is no greater than that under the pointwise constraint. Maximizing both quantities over \(\mu\) gives
\[
\IC^D(\IP_m,\varepsilon)\le\IC(\IP_m,\varepsilon).
\]
Together, these inequalities prove \eqref{eq:prior-free-bounds}. The stated bound on their difference follows from the common upper and lower bounds.
\end{proof}

\section{The Reciprocal-Rank Estimate}\label{sec:matrix-details}

\begin{proof}[Proof of Lemma~\ref{lem:reciprocal-rank}]

Let \(\widetilde{\boldsymbol\Lambda}\) be uniform over all matrices in \(\F_q^{t\times s}\), including the zero matrix, and let
\[
N=t-\rank\widetilde{\boldsymbol\Lambda}
\]
be the dimension of the kernel of \(\widetilde{\boldsymbol\Lambda}^{\mathsf T}\). We first bound \(\E N\) by counting the vectors in this kernel.

The kernel has \(q^N\) elements and always contains the zero vector. For any fixed nonzero \(\boldsymbol v\in\F_q^t\), the condition \(\widetilde{\boldsymbol\Lambda}^{\mathsf T}\boldsymbol v=\boldsymbol0\) imposes one nontrivial linear constraint on each of the \(s\) independent uniform columns of \(\widetilde{\boldsymbol\Lambda}\). Each constraint holds with probability \(1/q\), so
\[
\Pr[\widetilde{\boldsymbol\Lambda}^{\mathsf T}\boldsymbol v=\boldsymbol0]=q^{-s}.
\]
Summing over the vectors in the kernel and taking expectations gives
\[
\E q^N
=\E\bigl|\ker(\widetilde{\boldsymbol\Lambda}^{\mathsf T})\bigr|
=1+(q^t-1)q^{-s}.
\]
Since \(q^N\ge1+(q-1)N\) for every nonnegative integer \(N\), it follows that
\[
\E N\le\frac{(q^t-1)q^{-s}}{q-1}.
\]

We now condition on \(\widetilde{\boldsymbol\Lambda}\ne\boldsymbol0\), obtaining the uniform nonzero matrix used in the definition of \(\kappa_{q,t,s}\). The zero matrix has the largest possible nullity, \(N=t\), so excluding it cannot increase the expected nullity. Thus,
\[
\E[N\mid\widetilde{\boldsymbol\Lambda}\ne\boldsymbol0]
\le\E N.
\]
For any nonzero matrix of rank \(d\ge1\),
\[
\frac td=1+\frac{t-d}{d}\le1+(t-d)=1+N.
\]
Averaging this inequality over nonzero matrices therefore gives
\[
\begin{aligned}
\kappa_{q,t,s}
&=\E\left[\frac{t}{\rank\widetilde{\boldsymbol\Lambda}}
\;\middle|\;\widetilde{\boldsymbol\Lambda}\ne\boldsymbol0\right]\\
&\le1+\E[N\mid\widetilde{\boldsymbol\Lambda}\ne\boldsymbol0]\\
&\le1+\frac{(q^t-1)q^{-s}}{q-1}\\
&\le1+\frac1{q-1}\le2,
\end{aligned}
\]
where the last line uses \(t\le s\) and \(q\ge2\). This proves \eqref{eq:kappa-upper}. In particular, the stronger bound shows that \(\kappa_{q,t,s}\) tends to one as \(q\) grows, uniformly over \(1\le t\le s\).
\end{proof}

\end{document}